\documentclass[12pt,a4paper]{article}
\usepackage{amsmath,amssymb,mathrsfs,framed,esint,slashed}
\usepackage[colorlinks]{hyperref}
\usepackage{color,comment}
\usepackage[all]{xy}
\usepackage[color=yellow]{todonotes}

\usepackage{braket}
\usepackage{bbm}
\usepackage{enumerate}
\usepackage{cancel}

\newtheorem{theorem}{Theorem}[section]

\newtheorem{remark}[theorem]{Remark}

\newcommand{\Comment}[1]{\vspace{5mm}\par
\framebox{\begin{minipage}[c]{.95 \textwidth} \tt\bfi #1
\end{minipage}}\vspace{5 mm}\par}

\newcommand{\rem}[1]{}

\newcommand{\de}{{\rm d}}

\newcommand{\bq}{{\boldsymbol{q}}}
\newcommand{\bv}{{\boldsymbol{v}}}
\newcommand{\bp}{\boldsymbol{p}}

\newcommand{\bm}{{\boldsymbol{m}}}

\newcommand{\bx}{{\boldsymbol{x}}}

\newcommand{\bomega}{\boldsymbol{\omega}}
\newcommand{\bOmega}{\boldsymbol{\Omega}}

\newcommand{\bsigma}{\boldsymbol{\sigma}}

\newcommand{\bgamma}{\boldsymbol{\gamma}}

\newcommand{{\bLambda}}{\boldsymbol{\Lambda}}
\newcommand{\bmu}{\boldsymbol{\mu}}
\newcommand{\bnu}{\boldsymbol{\nu}}
\newcommand{\bbeta}{\boldsymbol{\beta}}

\newcommand{\bn}{{\boldsymbol{n}}}

\newcommand{\bw}{{\boldsymbol{w}}}
\newcommand{\br}{{\boldsymbol{r}}}
\newcommand{\bs}{{\boldsymbol{s}}}

\newcommand{\by}{{\boldsymbol{y}}}
\newcommand{\bA}{{\boldsymbol{A}}}

\newcommand{\bB}{{\boldsymbol{B}}}
\newcommand{\bR}{{\boldsymbol{R}}}

\newcommand{\bu}{{\boldsymbol{u}}}

\newcommand{\bXi}{{\boldsymbol{\Xi}}}

\newcommand{\bfi}{\bfseries\itshape}

\newcommand{\beq}{\begin{equation}}
\newcommand{\eeq}{\end{equation}}
\newcommand{\bal}{\begin{align}}
\newcommand{\eal}{\end{align}}
\newcommand{\ben}{\begin{eqnarray}}
\newcommand{\een}{\end{eqnarray}}

\renewcommand{\contentsname}{}

\numberwithin{equation}{section}
\numberwithin{figure}{section}

\begin{document}

\title{Holonomy and vortex structures\\in quantum hydrodynamics
}
\author{Michael S. Foskett,
 Cesare Tronci
\smallskip
\\ 
\small
\it Department of Mathematics, University of Surrey, UK
}
\date{\tiny \vspace{-.85cm}}
\maketitle

\begin{abstract} 
\small
We consider a new geometric approach to Madelung's quantum hydrodynamics (QHD) based on the theory of gauge connections. In particular, our treatment comprises a constant curvature thereby endowing QHD with intrinsic non-zero holonomy. In the hydrodynamic context, this  leads to a fluid velocity which no longer is constrained to be irrotational and   allows instead for  vortex filaments solutions. 
After exploiting the Rasetti-Regge method to couple the Schr\"odinger equation to vortex filament dynamics, the latter is then considered as a source of  geometric phase in the context of Born-Oppenheimer molecular dynamics. 
Similarly, we consider the Pauli equation for the motion of spin particles in electromagnetic fields and we exploit its underlying hydrodynamic picture to include vortex dynamics.

\end{abstract}

\vspace{-1.2cm}

{
\contentsname
\tableofcontents
}

\addtocontents{toc}{\protect\setcounter{tocdepth}{2}}

\newpage
\section{Introduction}\label{Intro-sec}

\subsection{The role of geometric phases}

In his seminal work \cite{Berry1984}, Berry was discovered that after undergoing an adiabatic cyclic evolution, a quantum system attains an additional phase factor independent of the dynamics and depending solely on the geometry of the evolution, since referred to as {\it Berry's phase}. This discovery opened up an entire field of study of the more general concept of {\it geometric phase} which has since been found to comprise the underlying mechanism behind a wide variety of physical phenomena, in both the classical and quantum domains. A key example of each is the Pancharatnam phase in classical optics, \cite{Pancharatnam1956}, which has been experimentally verified using experiments involving laser interferometry \cite{BhandariSamuel1988} and the celebrated Aharonov-Bohm effect of quantum mechanics, discovered in 1959 \cite{AharonovBohm1959} and experimentally verified in the late 80s \cite{TonomuraEtAl1986}, which was given a geometric phase interpretation in \cite{Berry1984}. One particular discipline which has benefitted greatly from the understanding of geometric phase is quantum chemistry, in which the separation of the molecular wavefunction into nuclear and electronic components gives rise to geometric phase effects in an array of phenomena \cite{Mead1992}, perhaps most famously including the Jahn-Teller effect \cite{JahnTeller1937, LonguetHigginsEtAl1958}.
To this day there is extensive research into the role of the geometric phase in quantum systems and in particular the application to molecular dynamics in quantum chemistry \cite{RyabinkinIzmaylov2013,RyabinkinEtAl2017,RyabinkinEtAl2014,GheribEtAl2015,AbediEtAl2012,RequistEtAl2016,AgostiniCurchod2018,Mead1992}.

Originally considered by Simon \cite{Simon1983}, the geometric interpretation of Berry's phase hinges on the gauge theory of principal bundles. The latter serves as a unifying mathematical framework so that  the geometric phase identifies the {\it holonomy} associated to the choice of {\it connection} on the bundle. In this picture one considers a base manifold $M$ with fibers isomorphic to a Lie group $G$ which can be put together in such a way to create a globally non-trivial topology.  The external parameters provide the coordinates on the base space, whilst the fibers are given simply by the $U(1)$ phase factor (not the phase itself!) of the wavefunction. Then, when one considers an adiabatic cyclic evolution of the external parameters, forming a closed loop in the base manifold, the corresponding curve in the $U(1)-$bundle (specified uniquely by a choice of {connection}) need not form a closed loop, with the extent of the failure to do so called the holonomy or Berry phase. This geometric picture on a principal bundle serves as the setting for all such geometric phase effects, with the base and fibers given by the problem under consideration. As mentioned before, holonomy manifests also in many classical systems, for example that discovered by Hannay \cite{Hannay1985}, with perhaps the most famous physical example being the Foucault pendulum \cite{WilczekShapere1989}. Such classical examples have led to much further study of these ideas, for example using reduction theory and geometric mechanics \cite{MarsdenEtAl1990} as well as applications to the $n$-body problem \cite{LittlejohnReinsch1997} and guiding center motion \cite{Littlejohn1988}.

Recently, a gauge theoretical description of quantum hydrodynamics in terms of connections has been suggested \cite{Spera2016}, using the Madelung transform \cite{Madelung1926,Madelung1927} to write the wavefunction in exponential form of an amplitude-phase product, $\psi = Re^{i\hbar^{-1}S}$. This change of variables has the well-known effect of transforming the Schr\"odinger equation into a hydrodynamical system upon defining a fluid velocity through the relation $\bv=m^{-1}\nabla S$. Whilst the hydrodynamic picture of quantum mechanics dates back to Madelung \cite{Madelung1926,Madelung1927}, it was Takabayasi \cite{Takabayasi1952, Takabayasi1983AB} who first realized that the circulation of the fluid velocity $\oint \bv\cdot\text{d}\bx$ must be quantized to ensure that the total wavefunction is single-valued, a fact which was later rediscovered by Wallstrom \cite{Wallstrom1994}. As shown in \cite{Spera2016}, the Madelung transform naturally allows one to consider a principal $U(1)$-bundle over $\mathbb{R}^3$ associated to the phase of the wavefunction. In this picture, the quantization condition of the circulation is in fact another example of holonomy, now corresponding to the connection ${\rm d}S$. More specifically, as the curvature of the connection vanishes everywhere except at those points for which $S$ is not single-valued, this is in fact a type of {\it monodromy}, with the exact value depending on the winding number of the loop surrounding the singularity.

Here we consider an alternative approach to holonomy in QHD, by using the Euler-Poincar{\'e} framework \cite{HolmMarsdenRatiu1998} to introduce a non-flat differentiable $U(1)$ connection whose constant curvature can be set as an initial condition. This results in holonomy with trivial monodromy as well as corresponding to a non-zero vorticity in {the hydrodynamic setting}. The key feature of this new approach to QHD is that it allows us to include geometric phase effects without entertaining double-valued functions or singular connections. Indeed, while the latter are still allowed as special cases, thereby leading to quantum vortex structures \cite{Takabayasi1952,Takabayasi1983,Takabayasi1983AB,BialynickiBirula1971,BialynickiBirulaSliwa2000}, here we shall apply the present construction to incorporate also non-quantized hydrodynamic vortex filaments, which are then coupled to the equations of quantum hydrodynamics. In this way, we provide an alternative approach to capture  geometric Berry phases  in the Born-Oppenheimer approximation or in the Aharonov-Bohm effect. We then consider the applications of this new approach to adiabatic molecular dynamics as well as extend the approach to include particles with spin by considering the Pauli equation. Motivated by the latter we then present the framework for introducing our connection in non-Abelian systems.

The remainder of this Introduction is devoted to presenting a more detailed exposition of the necessary background material and mathematical formalism to set the scene within which we present our results. Section \ref{Sec:QuantumGeometry} commences by outlining the necessary geometric structures of quantum mechanics used throughout this paper, including the Dirac-Frenkel variational principle and Hamiltonian structure of quantum mechanics, before switching to the hydrodynamic picture by introducing the Madelung transform and demonstrating the Lie-Poisson structure of QHD by using momentum map techniques. 
 The following Section \ref{Sec:QHDholonomy} presents the geometric interpretation of Wallstrom's quantization condition in terms of the holonomy of the multi-valued phase connection. Finally, we conclude the Introduction with an outline of the rest of the paper in \ref{Sec:Outline}, presenting a summary of the results in each subsequent 
section.

\subsection{Hamiltonian approach to quantum hydrodynamics}\label{Sec:QuantumGeometry}
In this section we provide the conventional geometric setting for quantum mechanics and its hydrodynamic formulation. As customary in the standard quantum mechanics of pure states, we consider a vector $\psi(t)\in \mathscr{H}$ in a Hilbert space $\mathscr{H}$. Then, the Schr{\"o}dinger equation $i\hbar\dot\psi=\widehat{H}\psi$ \cite{Schrodinger1926} can be derived from the Dirac-Frenkel (DF)  variational principle  \cite{FrenkelDirac1934}
\begin{align}
  0 = \delta\int_{t_1}^{t_2} \braket{\psi,i\hbar\dot\psi - 
  \widehat{H}\psi}\,\text{d}t\,,
  \label{DFVP}
\end{align}
where the bracket $\braket{\psi_1,\psi_2}$ denotes the real part of the Hermitian inner product $\braket{\psi_1|\psi_2}$. For the case of square-integrable wavefunctions $\mathscr{H}=L^2(\mathbb{R}^3)$, we have $\braket{\psi_1|\psi_2}=\int\!\psi_1^*\psi_2\,\de^3x$. The variational principle \eqref{DFVP} can be 
generalized  upon suitable redefinition of the total energy 
$h(\psi)$, which in the standard case considered here is simply given as the expectation of the Hamiltonian operator $\widehat{H}$, that is
$h(\psi) = \braket{\psi|\widehat{H}\psi}$. In fact, the Schr{\"o}dinger equation also admits a 
canonical Hamiltonian structure. For an arbitrary Hamiltonian $h(\psi)$, the generalized Schr{\"o}dinger 
equation reads
\begin{align}
  \frac{\partial\psi}{\partial t} = -\frac{i}{2\hbar}\frac{\delta h}{\delta \psi} =: X_h(\psi)\label{GeometricSchrodinger}\,,
\end{align}
in which $X_h$ is the corresponding Hamiltonian vector field. It can be checked that $h(\psi)=\langle\psi | \widehat{H}\psi\rangle$ recovers the standard Schr{\"o}dinger evolution. Then, the Hamiltonian structure arises from the symplectic form 
\begin{align}
\Omega(\psi_1,\psi_2)= 
2\hbar\,\text{Im}\braket{\psi_1|\psi_2}\label{symplecticform}
\end{align}
on $\mathscr{H}$.
It can then be readily verified that the corresponding Poisson bracket returns 
\eqref{GeometricSchrodinger} upon considering the 
relation $\dot{f}=\Omega(X_f,X_h)$. In the standard interpretation of quantum mechanics, $\mathscr{H}=L^2(\mathbb{R}^3)$ and the wavefunction identifies the probability density $D=|\psi|^2$ which  evolves according to 
\begin{equation}
  \frac{\partial D}{\partial t}  =\frac{2}{\hbar}\operatorname{Im}(\psi^*\widehat{H}\psi)
  \,.
\end{equation}
In the case of spin-less particles, the Hamiltonian operator $\widehat{H}$ is constructed out of the canonical observables $\widehat{Q}=\bx$ and $\widehat{P}=-i\hbar\nabla$ satisfying $[\widehat{Q}_i,\widehat{P}_j]=i\hbar\delta_{ij}$, so that $\widehat{H}=\widehat{H}(\widehat{Q},\widehat{P})$. As we shall now show, for the particular case of the physical Hamiltonian $\widehat{H}=\widehat{P}^2/2m+V(\widehat{Q})$, an equivalent hydrodynamic formulation of the theory emerges by rewriting the wavefunction in its  polar form ({\it Madelung transform})
$
\psi(\bx,t)=\sqrt{D(\bx,t)}e^{i S(\bx,t)/\hbar}
$.
Upon performing the appropriate substitutions in the DF variational principle \eqref{DFVP}, the latter becomes
   \begin{align}\label{DFLagrangian}
  0 = \delta\int_{t_1}^{t_2}\! \int\! D\left(\partial_tS +\frac{|\nabla S|^2}{2m} + \frac{\hbar^2}{8m}\frac{|\nabla 
  D|^2}{D^2}+V\right)\text{d}^3x\,\de t\,,
\end{align}
leading to
\begin{align}
&\partial_t D + \text{div}\left(\frac{D\nabla S}{m}\right)=0\label{DSEqn1} \,,\\
   &\partial_t S+\frac{|\nabla S|^2}{2m}-\frac{\hbar^2}{2m}\frac{\Delta\sqrt{D}}{\sqrt{D}}+V  
   =0\label{DSEqn2}\,.
   \end{align}
The first equation is clearly the continuity equation for the probability density $D=|\psi|^2$. 
The second equation resembles the Hamilton-Jacobi equation of classical mechanics, albeit with an additional 
term,
  often referred to as the {\it quantum potential}
\begin{align}\label{BohmPot}
  V_Q := -\frac{\hbar^2}{2m}\frac{\Delta\sqrt{D}}{\sqrt{D}} \,. 
\end{align}
Madelung's insight was to recognize that, written in terms of the variables $(D,S)$, the equation for conservation of probability takes the form of a fluid continuity equation for a fluid velocity field $\bv:=m^{-1}\nabla 
S$. Then, upon taking the gradient of the quantum Hamilton-Jacobi equation, both equations can be rewritten in terms of the new variables $(D,\bv)$ and 
one obtains the hydrodynamical system
\begin{align}
\partial_t D + \text{div}\left(D\bv\right)&=0 \label{QHD1}\,,\\
 m (\partial_t+\bv\cdot\nabla)\bv&= -\nabla(V_Q+V)\label{QHD2}\,,
   \end{align}
   corresponding to the {standard Eulerian form of hydrodynamic equations of motion,}
from here on referred to as the quantum hydrodynamic (QHD) equations. It is important to notice that as $\bv$ is a potential flow, the vorticity $\bomega := \nabla\times\bv$ of the flow is identically zero unless there are points where $S$ is multiple-valued.

   These quantum hydrodynamic equations were the starting point for 
   several further interpretations of quantum mechanics, most notably 
   Bohmian mechanics \cite{Bohm1952}, inspired by earlier works of de Broglie \cite{deBroglie1927}, in which one considers a physical particle that is guided by the quantum fluid. Whilst 
   we do not consider the implications of such theories here, we remark that 
   trajectory-based descriptions of quantum mechanics have been used to simulate 
      a wide variety of physical processes \cite{Wyatt2005} and the use of a QHD approach continues to be an active area of research, particularly in the field of quantum chemistry \cite{HoRaTr2021}. {More recently the geometric approach to QHD was extended to comprise} hybrid quantum-classical dynamics \cite{GayBalmazTronci2019}. With this in mind we shall consider the application of the fundamental ideas presented in this paper to quantum chemistry in Section \ref{Sec:BornOppenheimer}. 
   
We now turn our attention to the geometric structure underlying  QHD \cite{FoHoTr19,Fusca2015,KhesinEtAl2018}. It has  long been known that the map $\psi\mapsto (mD\bv,D)=:(\bmu,D)$ is a momentum map for the natural symplectic action of the semidirect-product group $\operatorname{Diff}(\mathbb{R}^3)\,\circledS\,\mathcal{F}(\mathbb{R}^3,U(1))$ on the representation space $L^2(\mathbb{R}^3)\simeq\operatorname{Den}^{1/2}(\mathbb{R}^3)$. Here, $\operatorname{Diff}(\mathbb{R}^3)$ denotes diffeomorphisms of physical space, $\mathcal{F}(\mathbb{R}^3,U(1))$ denotes the space of $U(1)-$valued scalar functions, while $\operatorname{Den}^{1/2}(\mathbb{R}^3)$ denotes the space of half-densities on physical space \cite{BatesWeinstein1997lectures}. Then, as described in \cite{Fusca2015,FoHoTr19}, it can be 
 shown that, for the case of the physical Hamiltonian operator, the total energy $h(\psi)=\text{Re}\int \psi^*\widehat{H}\psi\,\text{d}^3x$ 
can be expressed as a functional of these momentum map variables $(\bmu, D)$ to read
\begin{align}\label{QHDHam1}
h(\bmu, D)= \int\left[ \frac{|\bmu|^2}{2mD}+\frac{\hbar^2}{8m}\frac{|\nabla 
D|^2}{D}+DV(\boldsymbol{x})\right]\de^3x\,.
\end{align}
This process of expressing a Hamiltonian functional in terms of momentum map variables is known as  {\it Guillemin-Sternberg collectivization} \cite{Guillemin} and it leads to a Lie-Poisson system which in this case is defined on the dual of the semidirect-product Lie algebra $\mathfrak{X}(\mathbb{R}^3)\,\circledS\,\mathcal{F}(\mathbb{R}^3)$, where $\mathfrak{X}(\mathbb{R}^3)$ denotes vector fields on physical space while $\mathcal{F}(\mathbb{R}^3)$ stands for scalar functions. More explicitly, the Lie-Poisson bracket arising from the Madelung momentum map is as follows:
\begin{multline}\label{LPB1}
  \{f,g\}(\bmu, D) = \int \bmu \cdot \left(\left(\frac{\delta g}{\delta \bmu}\cdot\nabla\right)\frac{\delta f}{\delta \bmu} - \left(\frac{\delta f}{\delta \bmu}\cdot\nabla\right)\frac{\delta g}{\delta 
  \bmu}\right)\,\text{d}^3x\\
  + \int D\left(\left(\frac{\delta g}{\delta \bmu}\cdot\nabla\right)\frac{\delta f}{\delta D} - \left(\frac{\delta f}{\delta \bmu}\cdot\nabla\right)\frac{\delta g}{\delta 
 D}\right)\,\text{d}^3x\,,
\end{multline}
which coincides with the Lie-Poisson structure for standard barotropic fluid dynamics \cite{HolmSchmahStoica2009}. A vector calculus exercise \cite{Fusca2015} shows that Hamilton's equations $\dot{f}=\{f,h\}$ recover the hydrodynamic equations \eqref{QHD1}-\eqref{QHD2}. 

Despite its deep geometric footing, this construction invokes a Lie-Poisson reduction process \cite{HolmSchmahStoica2009,MarsdenRatiu2013} involving the existence of smooth invertible Lagrangian fluid paths $\eta\in\operatorname{Diff}(\mathbb{R}^3)$, so that $\dot{\boldsymbol\eta}(\bx_0,t)=\delta h/\delta\bmu|_{\bx=\boldsymbol\eta(\bx_0,t)}$. Thus, this description assumes that the phase $S$ is single-valued thereby leading to zero circulation and vorticity. In other words, this geometric formulation of QHD does not capture holonomy. As we shall see in the remainder of this paper, holonomy can still be restored by purely geometric arguments which however involve a different perspective from the one treated here.

\subsection{Multi-valued phases in quantum hydrodynamics}\label{Sec:QHDholonomy}

 Having previously investigated the geometry of QHD in terms of momentum maps \cite{FoHoTr19}, the rest of this paper focuses on an alternative geometric interpretation of QHD in terms of gauge connections. To see their role in QHD, we notice that whilst the equations of motion \eqref{QHD1}, \eqref{QHD2} had been known since the early days of quantum mechanics, in 1994 Wallstrom \cite{Wallstrom1994} 
   demonstrated that the requirement that the wavefunction be single-valued 
   does not actually imply that the QHD equations are  equivalent 
   to the Schr\"odinger equation   $i\hbar\partial_t\psi = -({\hbar^2}/{2m})\Delta\psi + 
  V\psi$, unless one has the 
   following additional condition on the phase of the wavefunction
   \begin{align}
  \oint_{c_0}\nabla S\cdot{\rm d}\bx = 2\pi\hbar n\label{QuantizedCirculation}\,, 
\end{align}
around any closed loop $c_0:[0,1]\to\mathbb{R}^3$ and for $n\in\mathbb{Z}$. This condition, equivalently asserting that the circulation of the fluid flow $\Gamma = \oint_{c_0}\bv\cdot{\rm d}\bx$ is quantized, is originally due to Takabayasi \cite{Takabayasi1952}, 
and arises due to the fact that $S$ can be considered a multi-valued function, by which we mean that the replacement
\begin{align}
  S(\bx)\to S(\bx) + 2\pi\hbar n 
\end{align}
leaves the wavefunction invariant. Then, the condition \eqref{QuantizedCirculation} is non-trivial, that is $n\neq 0$, whenever the curve $c_0$ encloses regions in which $S$ is multi-valued, {which 
itself occurs only at points where the wavefunction vanishes (nodes) \cite{GriffinKan1976,HirschfelderEtAl1974}. This can easily be seen by inverting the Madelung transform so that
\begin{align}
  S = -i\hbar\ln\left(\frac{\psi}{|\psi|}\right)  = \hbar\left(\arctan\left(\frac{{\rm Re}(\psi)}{{\rm Im}(\psi)}\right)+ 
  n\pi\right)\,,\label{MultivaluedS}
\end{align} 
where $n\in \mathbb{Z}$ arises from the multi-valued nature of the inverse 
tangent function}. In the hydrodynamic context, examples of this are given by the presence of vortices, 
a topic which has been studied extensively \cite{Takabayasi1952, Takabayasi1983, Takabayasi1983AB, BialynickiBirula1971, BialynickiBirulaSliwa2000} and which we will turn our attention to in Section \ref{Sec:Vortices}.

In fact, the condition \eqref{QuantizedCirculation} can be interpreted geometrically as in  \cite{Spera2016} by considering the 1-form $\nabla S\cdot{\rm d}\bx = {\rm d}S$ as a connection on a  $U(1)$-bundle over $\mathbb{R}^3$.
This is explained as follows. {Firstly, in writing $\psi=\sqrt{D}e^{iS/\hbar}$, we effectively make the decomposition $\mathbb{C}=\mathbb{R}^+\times U(1)$}.  We then consider the principal $U(1)$-bundle over $\mathbb{R}^3$, where the object $i\hbar^{-1}{\rm d}S$ can be considered as a $\mathfrak{u}(1)$-valued connection 1-form. 
Furthermore, it can be shown that the exponential of the loop integral  in \eqref{QuantizedCirculation} is indeed an element of the holonomy group of this bundle. See Chapter II in the standard reference \cite{KobayashiNomizu1963} for details on holonomy and principal bundles and \cite{WilczekShapere1989,ChruscinskiJamiolkowski2012,BohmEtAl2003, WayThesis2008, BohmBoyaKendrick1991} for their appearance in mechanical systems. 
As the curvature of the connection vanishes everywhere except at those points where $S$ is multi-valued, the quantization of the circulation can be explained geometrically, simply as the presence of the winding number through monodromy. 
 Following the discovery of the Berry phase \cite{Berry1984}
and its geometric interpretation due to Simon \cite{Simon1983}, this concept connecting geometric phases and holonomy via monodromy  has been found to arise in a wide variety of situations in quantum mechanics, ranging from QHD to the Aharonov-Bohm effect, and molecular dynamics.  
In this paper, we 
shall present a new alternative approach to QHD that features holonomy without monodromy by introducing a non-flat 
connection. 


\subsection{Outline of paper}\label{Sec:Outline}

Section \ref{Sec:PhaseConnections} presents the key idea of this paper, providing an alternative framework for understanding holonomy in QHD. Starting with \ref{Sec:PhaseConnectionsSubSec:1}, we introduce a different method of writing the 
wavefunction as an amplitude-phase product, allowing us to introduce a phase 
connection and write the Lagrangian in terms of this new dynamical variable. In deriving the equations of motion, we allow for this connection to possess non-trivial curvature resulting in new terms in phase equation. In Section \ref{Sec:PhaseConnectionsSubSec:2} we move to the fluid picture and show how these new terms give rise to a non-trivial circulation theorem and demonstrate that this connection carries non-trivial holonomy without monodromy. In Section \ref{SchrodingerReconstruction} 
we reconstruct the Schr\"odinger equation from the QHD system and see how the 
non-trivial curvature of the connection appears through minimal coupling, whilst 
Section \ref{Sec:Vortices} demonstrates how the non-zero vorticity can 
sustain solutions corresponding to vortex filaments and presents a coupled system of vortex dynamics within the Schr\"odinger equation.

In Section \ref{Sec:BornOppenheimer} we consider the application of these techniques to the Born-Oppenheimer 
factorization of the molecular wavefunction in adiabatic quantum chemistry. We begin in Section \ref{Sec:BOSubsec:1} 
by presenting the standard approach used in the literature and derive the QHD 
version of the equations of motion from the standard exponential polar form 
applied to the nuclear factor. This section also presents a summary of the key 
simplifications often used in the nuclear equation to aid simulations. In Section \ref{Sec:BOSubsec:2} we simply apply the formalism developed in Section \ref{Sec:PhaseConnections} to the Born-Oppenheimer system and comment on how these novel features may provide alternative viewpoints on key problems, which in the usual case arise due to the multi-valued nature of the objects in question. 
We conclude the section by considering a modified approach which couples a classical nuclear 
trajectory to a hydrodynamic vortex, incorporating geometric phase 
effects.

In Section \ref{Sec:ExactFact} we apply our treatment to exact factorization (EF) systems, in which one considers a wavefunction with additional parametric dependence.
We commence in Section \ref{Subsec:Nonad} by considering the time-dependent generalization of the Born-Oppenheimer ansatz, to which the name EF was given in \cite{AbediEtAl2012}, and apply our treatment to this system whilst also demonstrating its variational and Hamiltonian structures.
In Section \ref{SubSec:2Level}, we consider the exact factorization in the special case in which the `electronic factor' is given by a two-level system, thus leading us to introduce the spin density vector and accordingly specialize the geometric structures. Finally, as a particular case, we consider the exact factorization of the Pauli spinor in Section \ref{Sec:Pauli} and show that this treatment endows the hydrodynamic form of the Pauli equation with our additional holonomy. This section concludes with the coupling of the Pauli equation to vortex filament dynamics.

The method of deriving the QHD phase connection implies that it has vanishing curvature. In Section \ref{Sec:Non-Abelian} we show this 
explicitly and continue to show that in fact any gauge connection (corresponding to any arbitrary Lie group $G$) introduced 
in this way must have zero curvature. In this more general (non-Abelian) case, we 
also demonstrate how, for mechanical systems that give rise to connections 
through this approach, this zero curvature relation can be relaxed at the level of the Lagrangian, instead with the connection allowed to possess constant curvature as an initial condition, exactly as in the Abelian case of QHD presented in Section \ref{Sec:PhaseConnections}. 
\\

\section{Phase factors in quantum hydrodynamics}\label{Sec:PhaseConnections}
Having reviewed the standard approach to quantum hydrodynamics, including the interpretation of the quantized circulation condition as monodromy on a principal bundle, in this section we present an alternative approach in which non-trivial holonomy is 
built-in as an initial condition through a new dynamical connection. The point of departure is that, since the exponential map in $U(1)$ is not one-to-one, we are motivated to look at phase factors as elementary objects rather than expressing them as exponentials of Lie algebra elements in $\mathfrak{u}(1)=i\mathbb{R}$. As we shall see, this simple step eventually leads to a new method for incorporating holonomy in quantum hydrodynamics.

\subsection{Hamiltonian approach to connection dynamics}\label{Sec:PhaseConnectionsSubSec:1}
Instead of using the standard polar decomposition of the wavefunction, here we begin by  writing 
\begin{align}
  \psi(\bx,t)=\sqrt{D(\bx,t)}\,\theta(\bx,t)\label{OurPolar}\,,
  \qquad\text{with}\quad
  \theta\in 
\mathcal{F}(\mathbb{R}^3,U(1))
\,.
\end{align}
By writing explicitly the $U(1)$ factor $\theta$  we avoid using the exponential map which is not injective and works only with single-valued functions. Furthermore, this expression for the wavefunction has the advantage of making the { (gauge) Lie group $U(1)$ appear explicitly}, allowing us to use the tools of geometric mechanics \cite{MarsdenRatiu2013}. The relation  $\theta^*=\theta^{-1}$ allows us to write the Dirac-Frenkel variational principle \eqref{DFLagrangian} 
as
  \begin{align}
  0 = \delta\int_{t_1}^{t_2}\! \int\!  \bigg[i\hbar D \theta^{-1}\partial_t\theta - \frac{\hbar^2}{2m}\Big(\big|\nabla \sqrt{D}\big|^2 + D|\nabla\theta|^2\Big) - DV\bigg]\text{d}^3x\,\de t \label{Lagrangian2}\,.
\end{align}
Now, we let the phase factor $\theta(\bx,t)$  evolve according to the standard $U(1)$ action: 
\begin{align}
  \theta(\bx,t)=\Theta(\bx,t)\theta_0(\bx)\label{ThetaEvo}\,,
    \qquad\text{with}\quad
    \Theta\in 
\mathcal{F}(\mathbb{R}^3,U(1))\,.
\end{align}
In turn, this allows us to rewrite the 
time derivative as
\begin{align}
  \partial_t\theta = (\partial_t\Theta\,\Theta^{-1})\theta := {\xi}\theta\label{thetadot}\,,
     \qquad\text{where}\quad
     {\xi}\in \mathcal{F}(\mathbb{R}^3,\mathfrak{u}(1))
\end{align}
so that ${\xi}(\bx,t)$ is a purely imaginary function. 
We can also further evaluate terms involving the gradient of $\theta$ by introducing the 
connection $\bnu$ thus:
\begin{align}
\begin{split}
   \nabla\theta &= \nabla\Theta\, \theta_0 + \Theta\nabla\theta_0
  = \nabla\Theta\,\Theta^{-1}\theta - \Theta{\bnu}_0\theta_0\\
  &= -(- \nabla\Theta\,\Theta^{-1} + {\bnu}_0)\theta =: -{\bnu}\theta
\end{split}\label{NuConnectionDefinition}
\end{align}
where we have ${\bnu}_0:=-\nabla\theta_0/\theta_0$
and
${\nu}=\bnu\cdot\de\bx\in \mathfrak{u}(1)\otimes\Omega^1(\mathbb{R}^3)$, the factor  
 $\Omega^1(\mathbb{R}^3)$ being the space of differential one-forms on $\mathbb{R}^3$. Such approaches to introducing gauge connections have been used in 
the geometric mechanics literature before, for example in the study of liquid 
crystal dynamics \cite{Gay-BalmazRatiu2009,Gay-BalmazTronci2010,Gay-BalmazRatiuTronci2012,Gay-BalmazRatiuTronci2013,Holm2002,Tronci2012}. 
Indeed, we present the general formulation for gauge connections in {continuum} mechanical systems 
in Section \ref{Sec:Non-Abelian}.
\begin{remark}[Trivial and non-trivial connections]\label{jack}
As discussed in the next section, the gauge connection introduced in this way must have zero curvature. This is shown by taking the curl of the relation $\nabla\theta=-{\bnu}\theta$. In the present approach, we are exploiting this zero curvature case in order to have a final form of the QHD Lagrangian. Once variations have been taken in Hamilton's principle, the equations will be allowed to hold also in the case of non-zero curvature. 
This is a common technique used in geometric mechanics to derive new Lagrangians, for 
example used in \cite{BonetLuzTronci2015} to generalize the Dirac-Frenkel Lagrangian to include mixed state dynamics as well as in the study complex fluids \cite{Holm2002,Gay-BalmazRatiuTronci2012,Gay-BalmazRatiuTronci2013}. Within the theory of quantum dynamics, a similar approach was also used by Dirac in \cite{Dirac1931}.
\end{remark}
In terms of our newly introduced variables $\xi$ and $\bnu$, our variational principle 
reads
\begin{align}
  0 = \delta\int_{t_1}^{t_2}\! \int\!  \bigg[i\hbar D\xi -\frac{\hbar^2}{2m}\Big(D|\bnu|^2 + \big|\nabla \sqrt{D}\big|^2 \Big) - DV\bigg]\text{d}^3x\,\de t \label{Lagrangian2.1}\,,
\end{align}
where we have set $|\bnu|^2=\bnu^*\cdot\bnu$.
At this stage noting that both ${\xi}$ and ${\bnu}$ 
are purely imaginary, we define their real counterparts 
\begin{align}
 \bar{ \xi}:={i\hbar}{\xi}\,,\qquad \bar{\bnu}:={i\hbar}{\bnu}\label{barvariables}\,,
\end{align}
so that we can rewrite the variational principle \eqref{Lagrangian2.1} as 
\begin{align}
  0 = \delta\int_{t_1}^{t_2}\! \int\! D \Big( \bar{\xi}-\frac{|\bar{\bnu}|^2}{2m} - \frac{\hbar^2}{8m}\frac{|\nabla 
  D|^2}{D^2}  - V\Big)\,\text{d}^3x\,\de t=: \delta\int_{t_1}^{t_2}\!  \ell (\bar\xi, \bar\bnu, D)\,\de t\label{Lagrangian3}\,.
\end{align}
After computing  $(\delta\bar\xi,\delta \bar\bnu) = (\partial_t\bar\eta, -\nabla\bar\eta)$ with arbitrary $\bar\eta:= i\hbar \delta\Theta\,\Theta^{-1}$, taking variations
with arbitrary $\delta D$ yields the following general equations of motion:
\begin{align}
   \partial_t\left(\frac{\delta \ell}{\delta \bar\xi}\right) - \text{div}\left(\frac{\delta \ell}{\delta 
  \bar\bnu}\right)&= 0\,,\qquad\ 
  \partial_t\bar\bnu +\nabla\bar\xi= 0\,, \label{NuEvo}
  \qquad\ 
  \frac{\delta \ell}{\delta 
  D}= 0\,.
\end{align}
More specifically, the first equation leads to 
the continuity equation
\begin{align}
  \partial_tD + \text{div}(m^{-1}D{\bar\bnu})=0 \label{transport}\,,
\end{align}
in which we notice how $\widetilde\bnu:=m^{-1}\bar{\bnu}$ plays the role of a fluid velocity. Next, the third in \eqref{NuEvo} 
becomes
\begin{align}
  \bar\xi = &\ \frac{|\bar\bnu|^2}{2m} +V_Q + V 
  \label{xiHamiltonJacobi1}\,,
\end{align}
where we recall the quantum potential \eqref{BohmPot}.
Then, the second in \eqref{NuEvo}  leads to
\begin{equation}
\partial_t\bar\bnu+\nabla\bigg(\frac{|\bar\bnu|^2}{2m} +V_Q + V \bigg)=0
\,,
\label{bnueq}
\end{equation}
which formally coincides with the gradient of \eqref{DSEqn2}.
\begin{remark}[Lie-Poisson structure I] The new QHD equations \eqref{transport} and \eqref{bnueq} comprise a Lie-Poisson bracket on the dual of the semidirect-product Lie algebra $\mathcal{F}(\mathbb{R}^3)\,\circledS\,\Omega^1(\mathbb{R}^3)$. Specifically, the Lie-Poisson bracket for equations \eqref{transport}-\eqref{bnueq} reads
\begin{equation}
\{f,h\}=\int\left(\frac{\delta h}{\delta\bar\bnu}\cdot\nabla\frac{\delta f}{\delta D}-\frac{\delta f}{\delta\bar\bnu}\cdot\nabla\frac{\delta h}{\delta D}\right)\,{\rm d}^3 x
\,,
\label{AffLPB}
\end{equation}
while the Hamiltonian is
\[
h(D,\bar\bnu)=\int\!D\bigg(\frac{|\bar{\bnu}|^2}{2m} + \frac{\hbar^2}{8m}\frac{|\nabla 
  D|^2}{D^2}  + V\bigg)\,\de^3x
  \,.
\]
 The bracket \eqref{AffLPB} arises from a Lie-Poisson reduction on the semidirect-product group 
 \\
 $\mathcal{F}(\mathbb{R}^3, U(1))$ $\circledS\,\Omega^1(\mathbb{R}^3,\mathfrak{u}(1))$. Here, $\Omega^1(\mathbb{R}^3,\mathfrak{u}(1))$ denotes the space of differential one-forms with values in $\mathfrak{u}(1)\simeq i\mathbb{R}$, while the semidirect-product structure is defined by the affine gauge action $\bnu\mapsto\bnu+\Theta^{-1}\nabla\Theta$, where $\Theta\in\mathcal{F}(\mathbb{R}^3, U(1))$ and $\bnu\in\Omega^1(\mathbb{R}^3,\mathfrak{u}(1))$. For further details on this type of affine Lie-Poisson reduction, see \cite{Gay-BalmazRatiu2009,Holm2002}.
\end{remark}
Before continuing we specify the three different manifestations of the $U(1)$ connection 
we use in this paper. Firstly, we introduced the $\mathfrak{u}(1)$-valued connection 
$\bnu$ via the relation $\nabla\theta=:-\bnu\theta$. Then, its real counterpart $\bar\bnu$
was introduced via $\bar\bnu:= i\hbar\bnu\in\Omega^1(\mathbb{R}^3)$ which, for  $\theta = e^{iS/\hbar}$, coincides with $\nabla S$. Finally, in anticipation of the next section, we define $\widetilde{\bnu}\in\Omega^1(\mathbb{R}^3)$ 
by performing a further division by the mass, $\widetilde{\bnu}:=i\hbar 
m^{-1}\bnu = m^{-1}\bar\bnu$. This, corresponding to $m^{-1}\nabla S $ in the standard approach, 
will serve as the fluid velocity in the QHD picture.

Note how the usual exponential form $\theta = e^{iS/\hbar}$ in \eqref{OurPolar}
{returns} $\bar\xi = -\partial_t S$ and $\bar\bnu = \nabla S$, thus transforming equation \eqref{xiHamiltonJacobi1} 
into the standard phase equation (quantum Hamilton-Jacobi equation) \eqref{DSEqn2} of QHD. However, in our approach $\bar\bnu$ is now allowed to have a nontrivial  curvature as can be seen from the curl of \eqref{bnueq}:
\begin{align}\label{jeremy}
  \partial_t(\nabla\times\bar\bnu)=0\,.
\end{align}
This relation demonstrates explicitly how, in view of Remark \ref{jack}, the curvature of the connection $\bar\bnu$ need not be trivial (unlike ordinary QHD) but is instead  preserved in time. 
This is one of the main upshots of the {present}  approach.  We will develop this observation in the next section.


\subsection{Hydrodynamic equations and  curvature}\label{Sec:PhaseConnectionsSubSec:2}
In order to reconcile the new QHD equations with the Madelung-Bohm  construction,  this section discusses the hydrodynamic form of \eqref{transport}-\eqref{bnueq} in terms of the velocity-like variable $\widetilde{\bnu}$. As already seen the continuity equation is naturally rewritten in terms of $\widetilde\bnu$ as
\begin{align}
  \partial_tD+\text{div}(D\widetilde{\bnu})=0\,.\label{BracketEqn1}
\end{align}
Next, we multiply \eqref{bnueq} by $m^{-1}$ and expand the gradient to obtain the hydrodynamic-type equation
\begin{align}
  m(\partial_t+\widetilde{\bnu}\cdot\nabla)\widetilde{\bnu} = -\widetilde{\bnu}\times(\nabla\times\bar\bnu) - 
\nabla(V+V_Q)\label{BracketEqn2}\,.
\end{align}
This equation again clearly demonstrates the importance of not utilizing the exponential form of the phase as we now have the additional Lorentz-force term $-\widetilde{\bnu}\times(\nabla\times\bar\bnu)$ which is  absent in the equation \eqref{QHD2} of standard quantum hydrodynamics. Indeed, one sees that this additional term vanishes exactly when $\widetilde{\bnu}$ is a pure gradient. In the Bohmian interpretation, the Lagrangian fluid paths (aka {\it Bohmian trajectories}) are introduced via the relation $\dot{\boldsymbol\eta}(\bx,t)=\widetilde{\bnu}({\boldsymbol\eta}(\bx,t),t)$, so that Bohmian trajectories obey the Lagrangian path equation
\begin{equation}
m\ddot{\boldsymbol\eta}=-\dot{{\boldsymbol\eta}}\times\nabla\times\bar\bnu- 
\nabla_{\!\mathbf{x}}(V+V_Q)|_{\mathbf{x}=\boldsymbol\eta(\bx,t)}.
\end{equation}
We observe that a non-zero curvature modifies the usual equation of Bohmian trajectories by  the emergence of a Lorentz-force term. Notice that this term persists in the semiclassical limit {typically obtained by} ignoring the quantum potential contributions.

 We continue by writing the fluid equation in terms of the Lie derivative and the sharp isomorphism induced by the Euclidean metric in the fluid kinetic energy term in \eqref{Lagrangian3}. We have
\begin{align}
  m(\partial_t+\pounds_{\widetilde{\bnu}^\sharp})\widetilde{\bnu} = -\widetilde{\bnu}^\sharp\times(\nabla\times\bar\bnu) +\frac{m}{2}\nabla|\widetilde{\bnu}|^2- 
\nabla(V+V_Q)\,.
\end{align}
Then,   we obtain Kelvin's  circulation theorem  in the form
\begin{align}
  \begin{split}
0=  \frac{\text{d}}{\text{d}t}\oint_{c(t)}\widetilde{\bnu}\cdot\text{d}\bx &+  \frac1m
  \oint_{c(t)}\widetilde{\bnu}^\sharp\times(\nabla\times\bar\bnu)\cdot\text{d}\bx
  = \frac{\text{d}}{\text{d}t}\oint_{c_0}\widetilde{\bnu}\cdot\text{d}\bx
  \label{CirculationTheorem}\,.
  \end{split}
\end{align}
Here, $c(t)$ is a loop moving with the fluid velocity $\widetilde{\bnu}^\sharp$ such that $c(0)=c_0$ and the last equality follows directly from \eqref{jeremy}. 
In terms of 
the geometry of principal bundles, the last equality tells us that the holonomy of the connection $\bnu$ must be constant in time. Since no singularities are involved and $\bnu$ is assumed to be differentiable, the last equality in \eqref{CirculationTheorem} is an example of nontrivial holonomy with trivial monodromy.

%

We conclude this section with a short {comment on} helicity conservation.
By taking the dot product of the second equation in \eqref{NuEvo} with  $\nabla\times\widetilde{\bnu}$, 
\begin{align*}
 \partial_t (\widetilde{\bnu}\cdot\nabla\times\widetilde{\bnu})&= 
 -m^{-1}\rm{div}(\xi\nabla\times\widetilde{\bnu})\,,
\end{align*}
so that hydrodynamic helicity is preserved in time:
\begin{align}
\frac{\text{d}}{\text{d}t}\int \widetilde{\bnu}\cdot(\nabla\times\widetilde{\bnu})\,{\rm d}^3{x} &= 0\,.
\end{align}

Now that the hydrodynamic Bohmian interpretation has been discussed, it is not yet clear how this construction is actually related to the original Schr\"odinger equation of quantum mechanics. This is the topic of the next section.

\subsection{Schr{\"o}dinger equation with 	holonomy\label{SchrodingerReconstruction}}
In this section, we discuss the relation between the new QHD framework given by equations  \eqref{transport}-\eqref{bnueq} and the original Schr\"odinger equation. As a preliminary step, we consider the Helmholtz decomposition of $\bar\bnu$, that is
\begin{align}
  \bar\bnu(\bx,t) = \nabla s(\bx,t)+\hbar\nabla\times\bbeta(\bx)\label{Helmholtz1}\,,
\end{align}
where $\bbeta$ is a constant function to ensure \eqref{jeremy}. Also, we have added the factor $\hbar$ and fixed the Coulomb gauge $\operatorname{div}\bbeta=0$ for later convenience. Notice that although $s$ appears exactly in the place that $S$ would in the standard Madelung transform from Section \ref{Sec:QuantumGeometry}, here we have used the lowercase letter to emphasize that in this case we consider $s$ as a single-valued function. 
The relation \eqref{Helmholtz1} is reminiscent of similar expressions for the Bohmian velocity $\widetilde{\bnu}$ already appearing in \cite{BohmVigier1954}, although in the latter case these were motivated by stochastic augmentations of standard quantum theory. 
One can verify that upon substituting \eqref{Helmholtz1} into 
the equations of motion \eqref{NuEvo} and \eqref{transport}, the latter become
\begin{align}
 \partial_t s + \frac{|\nabla s+\hbar\nabla\times\bbeta|^2 }{2m}+ 
  V+V_Q&=0
  \,,\label{QHJ+}\\
  \partial_t D + \text{div}\left(D\,\frac{\nabla s+\hbar\nabla\times\bbeta}{m}\right) &=0
  \label{Continuity+}
\end{align}
Here, we have discarded numerical integration factors in the first equation. We recognize that these correspond to the standard Madelung equations for a free elementary charge in a magnetic field  $\hbar\Delta\bbeta$.


We will further characterize the 
Helmholtz decomposition \eqref{Helmholtz1}  of the connection $\bnu$ in terms of its defining relation \eqref{NuConnectionDefinition}. In particular, after constructing the Lagrangian \eqref{Lagrangian3}, $\bnu$ is then only defined as the solution of $\partial_t\bnu=-\nabla\xi$ in \eqref{NuEvo}. Combining the latter with $\partial_t\theta=\xi\theta$ leads to $\partial_t({\bnu}+ \theta^{-1}\nabla\theta)=0$ so that direct integration yields
\begin{align}
  {\bnu}=-\frac{\nabla\theta}{\theta} +i{\bLambda}(\bx)\,,\label{NuHelmholtz}
\end{align}
for a constant real function ${\bLambda}(\bx)$. An immediate calculation then shows that $
\nabla\times(\nabla\theta/\theta)=0$. Then, upon moving to the real-valued variables \eqref{barvariables}, direct comparison to \eqref{Helmholtz1} yields
\begin{align}
  \nabla s = -{i\hbar}\frac{\nabla\theta}{\theta}\,,\qquad 
\nabla\times\bbeta= -{\bLambda}\label{pbetarelation}\,.
\end{align}

{Now that we have characterized the additional terms due to the presence of non-zero curvature in the Madelung 
equations, we can use the expressions above to reconstruct the quantum Schr\"odinger 
equation.}
As we shall see, this coincides with the equation for $\psi=\sqrt{D}e^{is/\hbar}$, as it arises from the new Madelung equations \eqref{QHJ+}-\eqref{Continuity+}. Introducing $R=\sqrt{D}$ in \eqref{OurPolar}, we compute
\begin{align}
  i\hbar\partial_t\psi  &= i\hbar(\partial_t R\,\theta + R\,\partial_t\theta) \nonumber\\
   &= \left[-\frac{i\hbar}{m}\left(\frac{\nabla R}{R}\cdot\bar\bnu\right)-\frac{i\hbar}{2m}\text{div}(\bar\bnu)+\frac{|\bar\bnu|^2}{2m} +V_Q\right]\psi 
+ V\psi\,,
\label{prelimSchr}
\end{align}
having used the continuity equation in \eqref{transport} to find $\partial_t R$ and 
\eqref{thetadot} with \eqref{barvariables} and \eqref{xiHamiltonJacobi1} to find 
$\partial_t\theta$.
We must still manipulate the kinetic energy term to express everything in terms of 
$\psi$. Before continuing we notice that \eqref{OurPolar} leads to 
\begin{align}
    \nabla \psi=(R^{-1}\nabla R+\theta^{-1}\nabla\theta)\psi\,,
\end{align}
so that, since $\theta^{-1}\nabla\theta$ is purely imaginary,
\begin{align}
  \frac{\nabla R}{R} = \frac{\text{Re}(\psi^*\nabla\psi)}{|\psi|^2}\,, 
  \qquad
    \frac{\nabla \theta}{\theta} = 
    \frac{i\text{Im}(\psi^*\nabla\psi)}{|\psi|^2}
    \,.
\end{align}
Using  \eqref{NuHelmholtz} now leads to
$
\bar\bnu =  {\hbar\text{Im}(\psi^*\nabla\psi)}/{|\psi|^2} -\hbar
{\bLambda}$,
so that the right-hand side of \eqref{prelimSchr} can now be entirely written in terms of $\psi$. As shown in Appendix \ref{LongCalc}, lengthy calculations yield
\begin{align*}
 \left[-\frac{i\hbar}{m}\left(\frac{\nabla R}{R}\cdot\bar\bnu\right)-\frac{i\hbar}{2m}\text{div}(\bar\bnu)+\frac{|\bar\bnu|^2}{2m} +V_Q\right]\psi
  &= -\frac{\hbar^2}{2m}\Delta\psi + 
  \frac{i\hbar^2}{m}{\bLambda}\cdot\nabla\psi+ 
  \frac{\hbar^2}{2m}|{\bLambda}|^2 \psi\,.
\end{align*}
Putting everything back together we obtain the  Schr{\"o}dinger equation associated to the modified Madelung equations \eqref{QHJ+}-\eqref{Continuity+}:
\begin{align}
    i\hbar\partial_t\psi &= \left[\frac{(-i\hbar\nabla-\hbar{\bLambda})^2}{2m}+V\right]\psi\label{SchrodingerReconstructEqn}\,.
\end{align}
Notice how $\hbar{\bLambda}$, corresponding to the constant curvature part of our $U(1)$ connection, appears in the place of a magnetic vector potential in the Schr{\"o}dinger equation in which $\hbar$ plays the role of a coupling constant. {The quantity $\hbar{\bLambda}$ has been called {\it internal  vector potential} in quantum chemistry  \cite{SuDe91} and its} role is to incorporate holonomic effects in quantum dynamics. A static version of equation \eqref{SchrodingerReconstructEqn} also appeared in Dirac's work on singular electromagnetic fields \cite{Dirac1931}.
\begin{remark}[Lie-Poisson structure II]
Going back to Madelung hydrodynamics, we notice that the introduction of the vector potential ${\bLambda}$ leads to writing the hydrodynamic equation \eqref{BracketEqn2} in the form
\[
  m(\partial_t+\widetilde{\bnu}\cdot\nabla)\widetilde{\bnu} = \hbar\widetilde{\bnu}\times\nabla\times{\bLambda} - 
\nabla(V+V_Q)
\,.
\]
In turn, upon recalling the density variable $D=R^2$ and by introducing the momentum variable $\boldsymbol\mu=mD\widetilde{\bnu}+ \hbar D \bLambda$, the equation above possesses the alternative Hamiltonian structure (in addition to \eqref{AffLPB}) given by the hydrodynamic Lie-Poisson bracket \eqref{LPB1}, although the Hamiltonian \eqref{QHDHam1} is now modified to
\[
h(\bmu, D)= \int\left[ \frac{|\bmu-\hbar D{\bLambda}|^2}{2mD}+\frac{\hbar^2}{8m}\frac{|\nabla 
D|^2}{D}+DV(\boldsymbol{x})\right]\de^3x\,.
\]
\end{remark}
At this point topological defects may be incorporated by allowing $\bLambda$ in \eqref{SchrodingerReconstructEqn} to satisfy 
\begin{equation}
\oint_{c_0}\bLambda\cdot\de\bx=2\pi n\,.
\label{quantcond}
\end{equation}  
In this case, the present construction reduces to the standard approach to multivalued wavefunctions in condensed matter theory; see e.g. \cite{Kleinert}. 
In the variational framework, the velocity field $\widetilde\bnu=\delta h/\delta \bmu$ (introduced by the reduced Legendre transform \cite{MarsdenRatiu2013}) becomes the fundamental variable, with the resulting hydrodynamic Lagrangian being
\begin{align}
  \ell(\widetilde\bnu,D)=\int \bigg[\frac{1}{2}mD |\widetilde\bnu|^2+ \hbar D\widetilde\bnu\cdot\bLambda-\frac{\hbar^2}{8m}\frac{|\nabla 
D|^2}{D}-DV\bigg]\,\de^3x\,. \label{AlternativeQHDLagrangian}
\end{align}
Thus, the treatment in this paper  may accommodate both geometric and topological features depending on the explicit expression of $\bLambda$. 

Motivated by the appearance of the phase connection as a minimal coupling term in the Schr\"odinger equation \eqref{SchrodingerReconstructEqn}, it may be useful to  include the effect of an external magnetic field on the quantum system within this new QHD framework. Then, in the case of a spinless unit charge, the Schr\"odinger equation with holonomy reads
\begin{align}\label{modAB}
    i\hbar\partial_t\psi &= \left[\frac{(-i\hbar\nabla- (\hbar{\bLambda}+\bA))^2}{2m}+V\right]\psi\,.
\end{align}
In this instance ${\bLambda}$ and $\bA$ are  formally equivalent $U(1)$ gauge connections. Setting  ${\bLambda}=0$ in the Hamiltonian operator of \eqref{modAB} yields the Aharonov-Bohm Hamiltonian, in which case again the magnetic potential has a topological singularity.  However, despite the apparent equivalence between ${\bLambda}$ and $\bA$, they are related to essentially different features: while the holonomy associated to $\bA$ is associated to the properties of the external magnetic field, the holonomy associated to ${\bLambda}$ is intrinsically related to the evolution of the quantum state $\psi$ {itself}. This specific difference is particularly manifest in the case of two spinless unit charges moving within an external magnetic field. Indeed, in that case the 2-particle wavefunction $\psi(\bx_1,\bx_2,t)$ leads to defining  $\bLambda(\bx_1,\bx_2)$ and $\bA(\bx)$ on different spaces, thereby revealing their essentially different nature. At present, we do not know if this difference plays any role in the two-particle Aharonov-Bohm effect \cite{Samuelsson}, although we plan to develop this aspect in future work.


 \subsection{Schr\"odinger equation with hydrodynamic vortices}\label{Sec:Vortices}
 In this section, we show how the present setting can be used to capture the presence of vortices in quantum hydrodynamics. While the presence of topological vortex singularities in quantum mechanics has been known since the early days, this problem was considered in the context of the Madelung-Bohm formulation by Takabayasi  \cite{Takabayasi1952,Takabayasi1983, Takabayasi1983AB} and later by Białynicki-Birula  \cite{BialynickiBirula1971,BialynickiBirulaSliwa2000}. In the hydrodynamic context, the vorticity two-form $\omega$ is the differential of the Eulerian velocity field so that in our case $\bomega:=\nabla\times\widetilde{\bnu}$. Then, upon using \eqref{NuHelmholtz}, one has
 \begin{equation}\label{vorticity}
  \bomega(\bx) = -\frac{\hbar}{m}\nabla\times\bLambda(\bx)\,.
\end{equation}

In this section we wish to introduce the presence of hydrodynamic vortices in Schr\"odinger quantum mechanics. To this purpose,
we consider a hydrodynamic vortex filament of the form 
\begin{equation}
   \bomega(\bx) =\Gamma\int \bR_{\sigma}\,\delta(\bx-\bR(\sigma))\,\text{d}\sigma\,\label{vortexfilament}
\end{equation}
where $\bR(\sigma)$ is the curve specifying the filament, $\sigma$ is a 
parameterization of the curve and $\bR_{\sigma}:=\partial\bR/\partial\sigma$. Also, the number $\Gamma$ is the vortex strength and the expression $\Gamma=2n\pi\hbar/m$ recovers the particular case of quantized vortices \cite{Takabayasi1983AB}. For the quantization of three-dimensional vortex filaments, we refer the reader to \cite{GoMeSh}. Here,  in order to avoid problems with boundary conditions, we consider the simple case of vortex rings.
Then, inverting the curl operator in \eqref{vorticity} by using
the Biot-Savart law    \cite{Saffman1992} yields
\begin{align}\label{p-vort}
    \bLambda(\bx) = 
   (m/\hbar)\nabla\times\Delta^{-1}\bomega
 =
  \frac{m\Gamma}{\hbar}\nabla\times \int \!\bR_\sigma\, G(\bx-\bR)\,\de \sigma
\end{align}
where $G(\bx-\by)=-|\bx-\by|^{-1}/(4\pi)$ is the convolution kernel for the inverse Laplace operator $\Delta^{-1}$ and we have made use of \eqref{vortexfilament}.
Thus, the Schr\"odinger equation
\eqref{SchrodingerReconstructEqn} becomes
\begin{align}
    i\hbar\partial_t\psi &= \frac{1}{2m}\left(-i\hbar\nabla+{m\Gamma}\,\nabla\times \int \!\bR_\sigma\, G(\bx-\bR)\,\de \sigma\right)^{\!2}\psi+V\psi\label{SchrodingerVortices}\,,
\end{align}
where the vortex  position appears explicitly. In what follows, we will set $\Gamma=1$ for simplicity without affecting the general treatment.

Since in the present treatment the vorticity is constant, including the motion of the vortex filament in this description requires the addition of extra features. In quantum mechanics, the dynamics of hydrodynamic vortices has been given a Hamiltonian formulation by Rasetti and Regge \cite{RasettiRegge1975} and it was later developed in \cite{Holm03, PennaSpera1989, PennaSpera1992, KuznetsovRuban1998, KuznetsovRuban2000,Volovik2006}.
Then, one can think of exploiting the Rasetti-Regge approach to let the quantum vortex move {while interacting with the quantum state obeying} \eqref{SchrodingerVortices}. At the level of Hamilton's variational principle, this method leads to the following modification of the Dirac-Frenkel  Lagrangian in \eqref{DFVP}:
\begin{multline}
L = \frac{1}{3}\int\partial_t\bR\cdot\bR \times\bR_{\sigma} \,{\rm d}\sigma+\operatorname{Re}\int \!i\hbar\psi^*\partial_t\psi
\\ - \psi^* \bigg[\frac1{2m}{\bigg(-i\hbar\nabla+{m}\nabla\times \int \!\bR_\sigma\, G(\bx-\bR)\,\de \sigma\bigg)^{\!2}}+V\bigg]\psi\,{\rm d}^3{x}  \,,
\end{multline}
where the expression $\psi^*[\cdots]\psi$ on the last row identifies the Hamiltonian functional $h(\bR,\psi)$ satisfying $ \bR_\sigma\cdot{\delta h}/{\delta \bR}=0$ (valid for any Hamiltonian of the form $h=h(\bomega)$ \cite{HolmStechmann2004}) and 
\[
\bR_\sigma\times\left(\bR_\sigma\times\frac{\partial\bR}{\partial t}-\frac{\delta h}{\delta \bR}\right)=0
\,.
\]
Then, one finds
\[
\frac{\delta h}{\delta \bR}=-\frac{m}{\hbar}\bR_\sigma\times\frac{\delta h}{\delta \bLambda}\bigg|_{\bx=\bR}
={\hbar}\bR_\sigma\times\mathbb{P}\left(\operatorname{Im}(\psi^*\nabla\psi)-|\psi|^2\bLambda\right)\!\big|_{\bx=\bR\,},
\]
where $\mathbb{P}=\mathrm{Id}-\nabla\Delta^{-1}\operatorname{div}$ is the Leray projection on the divergence-free part. The coupled system reads
\begin{align}
\partial_t{\bR}=&\,{\hbar}\,\mathbb{P}\left(\operatorname{Im}(\psi^*\nabla\psi)-|\psi|^2\bLambda\right)\!\big|_{\bx=\bR\,}
+\kappa\bR_\sigma\,,
\\
i\hbar\partial_t\psi=&\,\frac1{2m}{(-i\hbar\nabla-\hbar\bLambda)^{2}}\psi+V\psi\,.
\end{align}
where $\bLambda$ is given as in \eqref {p-vort} and $\kappa$ is an arbitrary quantity. 
We also have that in the presence of vortex filaments the holonomy around a fixed loop $c_0$ 
can be expressed as
\begin{align}
-\hbar\oint_{c_0}\bLambda\cdot{\rm d}\bx =  m \int_{S_0}\int\text{d}\sigma\,\delta(\bx-\bR(\sigma))\, \bR_{\sigma}\cdot{\rm 
 d}\boldsymbol{S}\,,
\end{align}
via Stokes' theorem, where $S_0$ is a surface whose boundary defines the loop $\partial S_0 =: 
c_0$.

The idea of vortices in quantum mechanics has potentially interesting applications in the field of quantum chemistry. {For example, in 
the Born-Oppenheimer approximation,} the curvature of the Berry connection is given by a delta function at the point of conical intersections  \cite{Kendrick2003}. Once more, as conical intersections are topological singularities, the vortex structures generated by a singular Berry connection are quantized. After reviewing the general setting of adiabatic molecular dynamics, the next section shows how the construction in this paper can be used to deal with geometric phases in the Born-Oppenheimer approximation.


\section{Born-Oppenheimer molecular dynamics}\label{Sec:BornOppenheimer}
Motivated by the importance of geometric phase effects in quantum chemistry \cite{BohmBoyaKendrick1991,Kendrick2003, RyabinkinIzmaylov2013,RyabinkinEtAl2017,RyabinkinEtAl2014,GheribEtAl2015}, this section applies the formalism outlined in Section \ref{Sec:PhaseConnections} to the field of adiabatic molecular dynamics.

We start our discussion by  considering the starting point for all quantum chemistry methods, the Born-Oppenheimer ansatz for the molecular wavefunction. As explained in, for example, \cite{Marx,EntropyReviewNonadiabatic}, the  molecular wavefunction $\Psi(\{\br\},\{\bx\},t)$ for a system composed of $N$ nuclei with coordinates $\br_i$ and $n$ electrons with coordinates $\bx_a$ is factorized in terms of a nuclear wavefunction $\Omega(\{\br\},t)$ and a time-independent electronic function $\phi(\{\bx\};\{\br\})$ depending parametrically on the nuclear coordinates $\{\br\}_{i=1\dots N}$. In the simple case of a single electron and nucleus, we have
$\Psi(\br,\bx,t)= \Omega(\br,t)\phi(\bx;\br)$, so that the normalizations of $\Psi$ and $\Omega$ enforce $\int|\phi(\bx;\br)|^2\,\de^3x=1$. Equivalently, upon making use of Dirac's notation, we denote $\ket{\phi(\br)}:=\phi(\bx;\br)$ and write
\begin{align}
  \Psi(t) &= \Omega(\br,t)\ket{\phi(\br)}\label{StandardBOAnsatz}.
\end{align} 
The partial normalization condition becomes  $\|\phi(\br)\|^2:=\braket{\phi(\br)|\phi(\br)}=1$ and the Hamiltonian operator for the system reads $\widehat{H}=-\hbar^2M^{-1} \Delta/2\ +\widehat{H}_e$. Here,  $M$ is the nuclear mass and all derivatives are over the nuclear coordinate $\br$. In addition we have that the electronic state is the fundamental eigenstate of the electronic Hamiltonian $\widehat{H}_e$, so that $\widehat{H}_e\ket{\phi(\br)}=E(\br)\ket{\phi(\br)}$. 
The motivation for this ansatz comes from the separation of molecular motion into fast and slow dynamics due to the large mass difference between that of the electron and nucleus, the idea for which goes back to the original work of Born and Oppenheimer \cite{BornOppenheimer1927}.

\subsection{Variational approach to adiabatic molecular dynamics}\label{Sec:BOSubsec:1}

After applying the factorization ansatz \eqref{StandardBOAnsatz} and following some further manipulation involving integration by parts, the total energy $h= \text{Re}\int \braket{\Psi|\widehat{H}\Psi}\,\text{d}^3r$ of the 
system reads
\begin{align}
   h(\Omega)    &= \int \left[\Omega^*\frac{(-i\hbar\nabla+ \boldsymbol{\cal A})^2}{2M}\Omega + |\Omega|^2\epsilon(\phi,\nabla\phi)\right]\,\text{d}^3r\,,
\end{align}
in which we have introduced the {\it Berry connection} \cite{Berry1984}
\begin{align}\label{BerryConn}
\boldsymbol{\cal A}(\br):=\braket{\phi|-i\hbar\nabla\phi}\in\Omega^1(\mathbb{R}^3)\,,
\end{align}
and  defined the 
{\it effective electronic potential}
\begin{align}
  \epsilon(\phi,\nabla\phi):=E+\frac{\hbar^2}{2M}\|\nabla\phi\|^2 
  - \frac{|\boldsymbol{\cal A}|^2}{2M}\label{effectivepotential}\,.
\end{align} 
{Here, we have used the eigenvalue equation of $\widehat{H}_e$ while the last two terms correspond to the trace of the so-called {\it quantum geometric tensor}; see \cite{ProvostVallee1980} for details}. The appearance of the Berry connection is a typical feature of the Born-Oppenheimer method, which is  well-known  
 to involve non-trivial Berry phase effects \cite{MeadTruhlar1979, Mead1992, BohmEtAl2003}.
In order to write the nuclear equation of motion, we use the Dirac-Frenkel 
Lagrangian $L = \int i\hbar\Omega^*\partial_t\Omega\,\text{d}^3r - h(\Omega)$ and move to a hydrodynamical description. In the standard approach, one writes the nuclear function in the  polar 
form $\Omega(\br,t)=\sqrt{D(\br,t)}e^{iS(\br,t)/\hbar}$. Then, the previous DF Lagrangian becomes
\begin{align}
\label{BO-Lagr}
L(D,S,\partial_t S)   &= \int D\left(\partial_t S + \frac{|\nabla S+\boldsymbol{\cal A}|^2}{2M} + \frac{\hbar^2}{8M}\frac{|\nabla D|^2}{D^2} + 
\epsilon (\phi,\nabla\phi)\right)\,\text{d}^3r\,.
\end{align}

We notice that the 
 Born-Oppenheimer system is formally equivalent to standard quantum mechanics in the presence of an external {electromagnetic} field. Indeed, the Berry connection $\boldsymbol{\cal A}$ plays the role of the magnetic vector potential and one has a scalar potential in the form of $\epsilon$. Hence, the standard interpretation is to think of the nuclei evolving in an effective magnetic field generated by the {electronic motion}. In what follows, we shall adopt Madelung's hydrodynamic picture although an alternative approach using Gaussian wavepackets \cite{Littlejohn1986, Heller1976} is reported in Appendix \ref{GaussianAppendix}. Here, we proceed by applying Hamilton's principle $\delta\int_{t_1}^{t_2} \!L\,\de t=0$ for arbitrary variations $\delta D$ and $\delta S$, which returns the Euler-Lagrange 
equations
\begin{align}\label{BOMadelung}
  \frac{\partial D}{\partial t}+\text{div}\left(D\,\frac{\nabla S 
  +\boldsymbol{\cal A}}{M}\,\right)&= 0
  \,,\qquad\qquad\ 
    \frac{\partial S}{\partial t} + \frac{|\nabla S+\boldsymbol{\cal A}|^2}{2M} + V_Q + \epsilon = 0\,,
\end{align}
as usual understood as a quantum Hamilton-Jacobi equation for the nuclear phase and a continuity equation for the nuclear density $|\Omega|^2 = D$. Next, we follow the standard  approach by introducing  $\bv=M^{-1}\nabla S$. 
Finally, we write the Madelung equations in hydrodynamic form in terms of the velocity $\bu:=\bv+M^{-1}\boldsymbol{\cal A}$:
\begin{align}
       \partial_t D+\text{div}\left( D\bu\right)&= 0\,,\\
      M(\partial_t+\bu\cdot\nabla)\bu &= - 
  \bu\times\boldsymbol{\cal B} - \nabla\left(\epsilon + 
  V_Q\right)\,,
\end{align}
where $\boldsymbol{\cal B}:=\nabla\times\boldsymbol{\cal A}$. Notice how in this frame, a Lorentz force becomes apparent. 

The last equations above capture the nuclear motion completely, however in the quantum chemistry literature there are a variety of further specializations that can be made to the nuclear equation of motion, all aiming to alleviate computational difficulty. In the remainder of this section we summarize most of them by considering their subsequent effects on the nuclear fluid equation.

\begin{enumerate}

 \item{\bf Second order coupling:} In the 
 quantum chemistry literature it is often the case that the second order  coupling term,  $\braket{\phi|\Delta\phi}$ is neglected on the grounds that it has a negligible effect on the nuclear dynamics \cite{TullyNonadiabaticDynamics,Martinez1997}. 
 As can be verified directly upon expanding the real part one has that $\|\nabla\phi\|^2= -\text{Re}\braket{\phi|\Delta\phi}$, and hence such an approximation transforms equation \eqref{effectivepotential} to
\[
  \epsilon(\phi,\nabla\phi):=E
  - \frac{|\boldsymbol{\cal A}|^2}{2M}\,,
\]
{
At this stage, one is left with the Lorentz force acting on the nuclei as well as the potential given by the sum of the new effective electronic energy and nuclear quantum potential.}

 \item{\bf Real electronic eigenstate:} {Next, we make the assumption that the electronic eigenstate $\phi(\br)$ is  real-valued}, which is valid when the electronic Hamiltonian is non-degenerate \cite{TullyNonadiabaticDynamics,BredtmannEtAl2015}. The immediate consequence of the reality of $\phi$ is that the Berry connection $\boldsymbol{\cal A}:=\braket{\phi|-i\hbar\nabla\phi}$ 
 vanishes since the electronic phase is spatially constant. 
 In this case the nuclear fluid equation becomes
 \begin{align}
   (\partial_t+\bu\cdot\nabla)\bu &=- M^{-1}\nabla\left(E + 
  V_Q\right)\,.
 \end{align}
Clearly we still have  the nuclear quantum potential as well as the potential energy surface capturing electron-nuclear coupling. 

 \item{\bf Quantum Potential:} As detailed in \cite{FoHoTr19}, the quantum potential can also cause difficulties in numerical simulations. If we also consider
 neglecting the quantum potential term $V_Q$, the nuclear hydrodynamic equation can be written in its simplest form:
 \begin{align}
    (\partial_t+\bu\cdot\nabla)\bu &=- M^{-1}\nabla E\,.
 \end{align}
The quantum potential is usually  neglected
 by taking the singular weak  limit $\hbar^2  \rightarrow 0$ of the Lagrangian \eqref{BO-Lagr}.
Then, upon considering the single particle solution $D(\br,t)=\delta(\br-\bq(t))$, the nuclear equation 
 is equivalent to Newton's second law $M\ddot\bq=-\nabla E$ for a conservative 
 force.  
 \end{enumerate}
 It is only after this extreme level of approximation, neglecting all quantum terms (involving $\hbar$), that one obtains a classical equation of motion for the nuclei, in which one considers the picture of a nucleus evolving on a single 
 potential energy surface \cite{EntropyReviewNonadiabatic,TullyNonadiabaticDynamics}.

Whilst in this section we have considered adiabatic dynamics in the hydrodynamic picture via the Madelung 
transform, one can also proceed with an alternative approach in which the nuclear wavefunction is modelled by a frozen Gaussian wavepacket. This idea is presented in Appendix \ref{GaussianAppendix}, where we demonstrate how employing Gaussian coherent states within the 
variational principle \eqref{BO-Lagr} provides an alternative approach to regularizing the singularities that are known to arise in Born-Oppenheimer systems.

\subsection{Holonomy, conical intersections, and vortex structures}\label{Sec:BOSubsec:2}
In the context of the Born-Oppenheimer approximation, our approach to holonomy  can be applied by writing the nuclear wavefunction as in \eqref{OurPolar} while leaving the electronic wavefunction unchanged. For a real-valued electronic wavefunction, the Berry connection then vanishes and the Madelung equations \eqref{BOMadelung} are replaced by
\begin{align}\label{ourMT1}
  &{\partial_t D}+\text{div}(D\widetilde\bnu)= 0
  \,,
\\
 &M (\partial_t+\widetilde{\bnu}\cdot\nabla)\widetilde{\bnu}   = 
\hbar  \widetilde\bnu\times\nabla\times\bLambda - \nabla\bigg(E+\frac{\hbar^2}{2M}\|\nabla\phi\|^2+  V_Q\bigg).
\label{ourMT2}
\end{align}
We notice that in the case when the gauge potential $\bLambda$ is singular, these equations correspond to those appearing in the Mead-Truhlar method of adiabatic molecular dynamics \cite{MeadTruhlar1979}. This is a method to deal with the geometric phase  arising from double-valued electronic wavefunctions produced by the presence of conical intersections \cite{Kendrick2003,Mead1992, BurghardtEtAl2006}. This is called `molecular Aharonov-Bohm effect' \cite{Mead1980}. As double-valued  wavefunctions pose relevant computational difficulties, the Mead-Truhlar method performs a gauge transformation to move the presence of singularities from the wavefunction to the Berry connection.

 To illustrate the setting, we first consider the electronic eigenvalue problem 
\begin{equation}\label{eignevectors}
\widehat{H}_e\ket{\phi_n(\br)}=E_n(\br)\ket{\phi_n(\br)}
\end{equation}
and in particular the possibility that the first two separate eigenvalues (known as potential energy surfaces in the chemistry literature) intersect for a given nuclear configuration $\br_0$, that is $E_0(\br_0)=E_1(\br_0)$. In the previous sections, the fundamental eigenvalue $E_0$ was simply denoted by $E$. It is well-known that such intersections of the energy surfaces often form the shape of a double cone and are therefore referred to as {\it conical intersections} in the quantum chemistry literature. 
The non-trivial Berry phase that arises in such situations corresponds to the fact that the real electronic wavefunction $\ket{\phi_0(\br)}$ (previously denoted simply by $\ket{\phi(\br)}$) is double-valued around the point of degeneracy. The 
Mead-Truhlar method exploits the invariance of the electronic eigenvalue problem under the gauge transformation 
$
\ket{\phi(\br)}\mapsto 
\ket{\phi'(\br)}=e^{i\zeta(\br)/\hbar}\ket{\phi(\br)}
$
 thereby redefining the Berry connection according to
$
 \boldsymbol{\cal A} \mapsto  \boldsymbol{\cal A}'=\boldsymbol{\cal A} +\nabla 
 \zeta$. 
 Specifically, one selects $\zeta$ such that the phase $e^{i\zeta/\hbar}$ exactly compensates the double-valuedness of $\ket{\phi}$ resulting in the new electronic state $\ket{\phi'}$ being 
single-valued \cite{Kendrick2003,Mead1992} and thus avoiding the need to deal with double-valued functions. However, since $\boldsymbol{\cal A}=0$ (because $\ket{\phi}$ is real) and $e^{i\zeta/\hbar}$ must be multi-valued, such a transformation has the cost that the corresponding vector potential $\boldsymbol{\cal A}' = \nabla\zeta$ is singular at the point of the conical intersection. Then, after  replacing $|\phi\rangle\to|\phi'\rangle$, one obtains the equations \eqref{ourMT1}-\eqref{ourMT2} in the case $\widetilde\bnu=M^{-1}\nabla S$ and $\hbar\bLambda=\nabla\zeta$, so that the problem under consideration becomes equivalent to the Aharonov-Bohm problem, whence the name `molecular Aharonov-Bohm effect'.

While conical intersections are essentially topological defects, the physical consistency of these singularities has been recently questioned by Gross and collaborators \cite{Min14,RequistEtAl2016}. In their work, it is argued that the emergence of these topological structures is intrinsically associated to the particular type of adiabatic model arising from the Born-Oppenheimer factorization ansatz. Indeed, the results in \cite{Min14,RequistEtAl2016} and following papers show that these type of singularities are absolutely absent in the exact case of nonadiabatic dynamics. This leads to the question of whether alternative approaches to adiabatic dynamics can be obtained in order to avoid dealing with conical intersections. Notice that the absence of these defects does not imply the absence of a geometric phase. Indeed, as the Berry connection is not generally vanishing in nonadiabatic dynamics, this leads to nontrivial holonomy which in turn does not arise from topological singularities. In this context, a gauge connection associated to hydrodynamic vortices as in Section \ref{Sec:Vortices} may be representative of an alternative molecular geometric effect in which the geometric phase depends on the integration loop. In this case, one could  drop the quantum potential in \eqref{ourMT2} and select the particle solution $D(\br,t)=\delta(\br-\bq(t))$ in \eqref{ourMT1}.
However, due to \eqref{vorticity} and \eqref{vortexfilament}, this approach would produce a $\delta-$like Lorentz force in the nuclear trajectory equation, thereby leading to major difficulties. 
The latter may be overcome by finding appropriate closures at the level of Hamilton's principle. For example, one could use Gaussian wavepackets as presented in Appendix \ref{GaussianAppendix}. However, here we adopt  a method  inspired by previous work in plasma physics \cite{Ho83} and geophysical fluid dynamics \cite{Ho86}. Let us start with the hydrodynamic Lagrangian, of the type \eqref{AlternativeQHDLagrangian}, underlying equations \eqref{ourMT1}-\eqref{ourMT2}:
$\ell(D,\widetilde\bnu)=\int\!D\left({M}|\widetilde\bnu|^2/2+\hbar\widetilde\bnu\cdot\bLambda-V_Q-\epsilon\right)\de^3 r$.
Here, the Eulerian variables $D(\br,t)$ and $\widetilde\bnu(\br,t)$ are related to the  Lagrangian fluid path $\boldsymbol\eta(\br,t)$ (Bohmian trajectory) by the relations $\dot{\boldsymbol\eta}(\br,t)=\widetilde\bnu({\boldsymbol\eta}(\br,t),t)$ and $D({\boldsymbol\eta}(\br,t),t)\,\de^3{\eta}(\br,t)= D_0(\br)\,\de^3r$. If we now restrict the Bohmian trajectory to be of the type ${\boldsymbol\eta}(\br,t)=\br+\bq(t)$, we have $\widetilde\bnu(\br,t)=\dot{\bq}(t)$ and $D(\br,t)=D_0(\br-\bq(t))$. Then, since in this case {$\int \! DV_Q\,\de^3 r=\text{const.}$}, the Lagrangian $\ell(D,\widetilde\bnu)$ becomes
\begin{align}
L(\bq,\dot\bq)=\frac{M}2|\dot\bq|^2+\int\!D_0(\br-\bq)\big[\hbar\dot\bq\cdot\bLambda(\br)-\epsilon(\br)\big]\,\de^3 r 
\label{Jeff}
\,.
\end{align}
Here, $D_0$ is typically a Gaussian distribution and we recall that we are considering the case of a real electronic wavefunction, so that $\boldsymbol{\cal A}=0$ in the definition of $\epsilon$ \eqref{effectivepotential}. Then, one obtains the  Euler-Lagrange equation
\begin{equation}\label{particles+vortices}
M\ddot\bq=\hbar\dot\bq\times\nabla\times\int \!D_0(\br-\bq)\bLambda(\br)\,\de^3 r
-  \nabla\!\int \!\epsilon(\br)D_0(\br-\bq)\,\de^3 r
\,,
\end{equation}
where $\bLambda$ is given as in \eqref{p-vort}. (This method  was recently applied to the Jahn-Teller problem by one of us in \cite{RaTr20}, where it was shown to recover some recent results obtained from  exact nonadiabatic treatments \cite{Gross3}.)  We see that the nuclear density acts as a convolution kernel regularizing both the connection $\bLambda$ and the potential energy surface appearing in $\epsilon$. For example, this method could be used to regularize topological singularities arising from conical intersections. Then, the geometric phase and the regularized potential energy surface read
\[
-\hbar\oint_{c_0}\int \!\de^3 r'\,D_0(\br'-\br)\bLambda(\br')\cdot\de\br
\,,\qquad\qquad 
\int\!\de^3 r'\,D_0(\br'-\br)E(\br')
\,.
\]
Also, in the present context, the self-consistent vortex evolution may be included upon constructing a Rasetti-Regge type Lagrangian by the replacement $L(\bq,\dot{\bq})\to L(\bq,\dot{\bq})+(1/3)\int\!\partial_t\bR\cdot\bR \times\bR_{\sigma} \,{\rm d}\sigma$, similarly to the approach in Section \ref{Sec:Vortices}. In this case, the vortex evolution equation reads $\partial_t\bR=MD_0(\bR-\bq)\dot\bq +\kappa\bR_\sigma$.


\section{Exact wavefunction factorization}\label{Sec:ExactFact}
\subsection{Nonadiabatic molecular dynamics}\label{Subsec:Nonad}

While the Born-Oppenheimer approximation has been very successful in the modeling of adiabatic molecular dynamics, many processes in both chemistry and physics involve  quantum electronic transitions. In this case, the Born-Oppenheimer approximation breaks down so that nonadiabatic processes require a new modeling framework. In this context, the molecular wavefunction $\Psi(\br,\bx,t)$ in \eqref{StandardBOAnsatz} is expanded in the basis provided by the spectral problem \eqref{eignevectors} so that the resulting series expansion is known as {\it Born-Huang expansion} \cite{BH56}. While this expansion is the basis for several ab initio methods in nonadiabatic molecular dynamics, an alternative picture, originally due to Hunter \cite{Hunter1975}, has recently been revived by Gross and collaborators \cite{AbediEtAl2012}. In this alternative picture, the exact solution of the molecular Schr\"odinger equation is written by allowing the electronic wavefunction in \eqref{StandardBOAnsatz} to depend explicitly on time, that is
\begin{align}
  \Psi(t) &= \Omega(\br,t)\ket{\phi(\br,t)}\label{EFAnsatz}.
\end{align} 
Going back to von Neumann \cite{vonNeumann}, this type of wavefunction factorization also represents the typical approach to Pauli's equation for charged particle dynamics in electromagnetic fields \cite{Takabayasi1955}. In this context, the electronic coordinates are replaced by the spin degree of freedom while the nuclear coordinates are replaced by the  particle position coordinate. 
In more generality, the exact factorization picture is applicable in a variety of multi-body problems in physics and chemistry. In this section, we investigate the role of holonomy in the exact factorization picture. Then, when the holonomy arises from a delta-like curvature, we shall present how the quantum hydrodynamics with spin is modified by the presence of vortex filaments. Before moving on to the Pauli equation, here we shall present the general hydrodynamic equations arising from the exact factorization \eqref{EFAnsatz}.

The typical Hamiltonian for the total system is written as 
\begin{equation}\label{totHam}
\widehat{H}_\textrm{tot}=-\frac{\hbar^2}{2M}\,\Delta+\widehat{H}(\br)
\end{equation}
so that, after rearrangement, replacing \eqref{EFAnsatz} in the Dirac-Frenkel Lagrangian 
\begin{align*}
L=\operatorname{Re}\int\!\braket{\Psi|(i\hbar\partial_t-\widehat{H}_\textrm{tot})\Psi}\,\de^3 r
\end{align*}
leads to the exact factorization Lagrangian
\begin{multline}\label{EFLagr}
L =\text{Re}\int\! i\hbar\big(\Omega^*\partial_t\Omega+|\Omega|^2\langle\phi|\partial_t\phi\rangle\big)\,\text{d}^3r
\\
    - \int \!\bigg[ \frac1{2M}\Omega^*(-i\hbar\nabla+\boldsymbol{\cal A})^2\Omega+|\Omega|^2\left(\langle\phi|\widehat{H}\phi\rangle+\frac{\hbar^2}{2M}\|\nabla\phi\|^2 - 
  \frac{|\boldsymbol{\mathcal{A}}|^2}{2M}\right)
    \bigg]\text{d}^3r\,,
\end{multline}
where the second line identifies (minus) the expression of the total energy and we recall \eqref{BerryConn}. Notice that we have not enforced the partial normalization condition $\|\phi(\br)\|^2=1$ in the variational principle, although this condition can be verified to hold a posteriori after taking variations.
When dealing with molecular problems, $|\phi\rangle=\phi(\bx,t;\br)$ is a time-dependent electronic wavefunction \cite{AbediEtAl2012} and $\widehat{H}(\br)$ is the electronic Hamiltonian in \eqref{eignevectors} from Born-Oppenheimer theory. On the other hand, if we deal with $1/2-$spin degrees of freedom, $|\phi\rangle=|\phi(\br,t)\rangle$ is a two-component complex vector parameterized by the coordinate $\br$, while $\widehat{H}(\br)$ is of the form $\widehat{H}(\br)=a(\br)+\boldsymbol{b}(\br)\cdot\widehat{\boldsymbol{\sigma}}$ and $\widehat{\boldsymbol{\sigma}}$ denotes the array of Pauli matrices $\widehat{\boldsymbol{\sigma}}=(\widehat{{\sigma}}_x,\widehat{{\sigma}}_y,\widehat{{\sigma}}_z)$.

At this point, the standard approach to QHD requires replacing the polar form of $\Omega$ in \eqref{EFLagr}. If instead we adopt the method developed in Section \ref{Sec:PhaseConnections} and write $\Omega$ as in \eqref{OurPolar}, we obtain the following hydrodynamic Lagrangian of Euler-Poincar\'e type \cite{HolmMarsdenRatiu1998}:
\begin{multline}
 \ell(D,\bar{\xi},\bar{\nu},|\phi\rangle)= \int \!
 D\bigg(\bar{\xi} - \frac{|\bar\bnu +\boldsymbol{\mathcal{A}}|^2}{2M} -\frac{\hbar^2}{8M}\frac{|\nabla D|^2}{D^2} + \Phi
 \\
 +\langle\phi|\widehat{H}\phi\rangle -\frac{\hbar^2}{2M}\|\nabla\phi\|^2 + 
  \frac{|\boldsymbol{\mathcal{A}}|^2}{2M}\bigg)\,\text{d}^3r\label{EFReducedLagrangian}\,,
\end{multline}
where $\Phi(\br) := \braket{\phi|i\hbar   \partial_t\phi}$. At this stage one 
can compute the corresponding equations of motion.

Upon introducing the hydrodynamic velocity $\bv$ and the matrix density $\tilde\rho$,
\begin{align}
\bv(\br,t)=M^{-1}(\bar\bnu+\boldsymbol{\mathcal{A}})
\,,\qquad\qquad
\tilde\rho(\br,t)=D|\phi\rangle\langle\phi|
\,, \label{philip}
\end{align}
we obtain the hydrodynamic equations in the form
\begin{align}
  \partial_tD+ \text{div}(D\bv)&= 0\,, \label{EF1}\\
 MD(\partial_t+\bv\cdot\nabla)\bv &= \hbar D\bv \times\nabla\times\bLambda +D\nabla V_Q- 
\braket{\tilde\rho|\nabla\widehat{H}} - 
  \frac{\hbar^2}{2M}\partial_j\left(D^{-1}\braket{\nabla\tilde\rho|\partial_j\tilde\rho}\right)\,,\label{EF2}\\
  i\hbar(\partial_t\tilde\rho + \text{div}(\tilde\rho\bv))&= \bigg[\widehat{H} - \frac{\hbar^2}{2M}\text{div}(D^{-1}\nabla\tilde\rho),\tilde\rho\bigg] 
\label{EF3}\,.
\end{align}
Here, we have used the notation 
\begin{align*}
    \langle A(\br)|B(\br)\rangle=\operatorname{Tr}(A(\br)^\dagger B(\br))
\end{align*}
where $\operatorname{Tr}$ denotes the matrix trace.

Interestingly, as shown  in \cite{FoHoTr19} for an equivalent variant of these equations, this system can also be derived via the standard Euler-Poincar{\'e} variational principle for hydrodynamics (as presented earlier in Section \ref{Sec:QuantumGeometry}) in which the Eulerian quantities $D(\br,t)$ and $\bv(\br,t)$ are related to the Lagrangian (Bohmian) trajectory $\boldsymbol\eta(\br,t)$ by the relations $\dot{\boldsymbol\eta}(\br,t)=\bv({\boldsymbol\eta}(\br,t),t)$ and $D({\boldsymbol\eta}(\br,t),t)\,\de^3{\eta}(\br,t)= D_0(\br)\,\de^3r$. In this instance, as seen before, equations 
\eqref{EF1}-\eqref{EF3} correspond to a Lagrangian of the type as given by 
\eqref{AlternativeQHDLagrangian}, here written as
\begin{align}\label{EFHDLagr}
  \ell(\bv, D, \Xi, \tilde\rho) = \int\! \left[\frac{1}{2}MD|\bv|^2 + \hbar D\bv\cdot\bLambda+\frac{\hbar^2}{8M}\frac{|\nabla D|^2}{D}+\braket{\tilde\rho, i\hbar\Xi - \widehat{H}} 
  -\frac{\hbar^2}{4M}\frac{\|\nabla\tilde\rho\|^2}{D}\right] \text{d}^3r\,,
\end{align}
in which $\Xi(\br,t)$ is defined by the Schr\"odinger equation  $\partial_tU(\br,t)=\Xi({\boldsymbol\eta}(\br,t),t)U(\br,t)$ for the propagator $U(\br,t)$ with {$\tilde\rho({\boldsymbol\eta}(\br,t),t)\,\de^3{\eta}(\br,t)= U(\br,t)\tilde{\rho}_0(\br)U^\dagger(\br,t)\,\de^3r$}. 
Then,  applying Hamilton's principle by combining the variations 
\begin{align}\label{vars}
  \delta D &= -\text{div}(D\bw)\,,\qquad \delta \bv = \partial_t\bw + 
(\bv\cdot\nabla)\bw - (\bw\cdot\nabla)\bv\,,\\
\delta\tilde\rho &= [{\Upsilon},\tilde\rho] - \text{div}(\widetilde\rho\bw)\,,\qquad \delta \Xi = \partial_t{\Upsilon} + [{\Upsilon},\Xi] - \bw\cdot\nabla\Xi + 
\bv\cdot\nabla{\Upsilon}\,,
\end{align}
with the auxiliary equations $\partial_tD + \text{div}(D\bv) = 0$ and $\partial_t\tilde\rho + \text{div}(\tilde\rho\bv) = [\Xi,\tilde\rho]$ returns the  equations \eqref{EF1}-\eqref{EF3}. Here, $\bw$ and $\Upsilon$ are both arbitrary and vanish at the endpoints. In the special case when $\bLambda=0$, a detailed presentation of this approach can be found in \cite{FoHoTr19}.
\begin{remark}[Semidirect product Lie-Poisson structure]
  Upon defining $\bm:=MD\bv+\hbar D\bLambda$, the equations \eqref{EF1}-\eqref{EF3} acquire the following Poisson structure
  \begin{align}\nonumber
\{f,g\}(\bm,D,\tilde\rho)=&\,\int\!\bm\cdot\left(\frac{\delta g}{\delta \bm}\cdot\nabla\frac{\delta f}{\delta \bm}-\frac{\delta f}{\delta \bm}\cdot\nabla\frac{\delta g}{\delta \bm}\right)\de ^3 r
\\
&\,
-
\int \!D\left(\frac{\delta f}{\delta \bm}\cdot\nabla\frac{\delta g}{\delta D}-\frac{\delta g}{\delta \bm}\cdot\nabla\frac{\delta f}{\delta D}\right)\de ^3 r
\nonumber
\\
&\,
{-\int \bigg\langle \tilde\rho,\,\frac{i}{\hbar}\bigg[\frac{\delta f}{\delta \tilde\rho},\frac{\delta g}{\delta \tilde\rho}\bigg]+\frac{\delta f}{\delta \bm}\cdot\nabla\frac{\delta g}{\delta \tilde\rho}-\frac{\delta g}{\delta \bm}\cdot\nabla\frac{\delta f}{\delta  \tilde\rho} \bigg\rangle\,\de ^3 
r\,,}
\label{LPB-EF}
\end{align}
which is accompanied by the Hamiltonian functional
\begin{align}
  h(\bm, D, \tilde\rho) =\int\!\bigg[\frac{|\bm-\hbar D\bLambda|^2}{2MD} -\frac{\hbar^2}{8M}\frac{|\nabla D|^2}{D}+ \braket{\tilde\rho, \widehat{H}} 
  +\frac{\hbar^2}{4M}\frac{\|\nabla\tilde\rho\|^2}{D}\bigg]\de^3r\,.
\end{align}
We remark here that the change of variables $\tilde\rho\to i\hbar\tilde\rho$ takes the bracket 
\eqref{LPB-EF} into a  Lie-Poisson bracket on the dual of the semidirect-product Lie algebra 
$\mathfrak{X}(\mathbb{R}^3)\,\circledS\, \big(\mathcal{F}(\mathbb{R}^3)\oplus\, \mathcal{F}\big(\mathbb{R}^3,\mathfrak{u}(\mathscr{H}_e)\big)\big)$. 
   \end{remark}

\subsection{Exact factorization for electronic two-level systems}\label{SubSec:2Level}
Instead of treating infinite-dimensional two-particle systems, the remainder of this paper studies the case in which the electronic state is represented by a two-level system so that
\[
\tilde\rho(\br,t)=\frac{D(\br,t)}{2}\left(\mathbbm{1}+\frac2\hbar{\boldsymbol{s}}(\br,t)\cdot\widehat{\boldsymbol{\sigma}}\right)=:\frac1{2}\left(D(\br,t)\mathbbm{1}+\frac2\hbar\tilde{\boldsymbol{s}}(\br,t)\cdot\widehat{\boldsymbol{\sigma}}\right)
,
\]
where $\bs$ is the {\it spin vector} as described in \cite{BohmSchillerTiomnoA1955,BohmSchillerB1955,Takabayasi1955,Bialynicki-Birula1995Weyl}, given by $\bs = \hbar\braket{\phi|\boldsymbol{\widehat{\sigma}}\phi}/2$ and satisfying $|{\boldsymbol{s}}|^2=\hbar^2 /4$, and $\mathbbm{1}$ 
is the $2\times 2$ identity operator. In addition, the Hamiltonian operator reads $\widehat{H}=a(\boldsymbol{r})\mathbbm{1}+\boldsymbol{b}(\boldsymbol{r})\cdot\widehat{\boldsymbol{\sigma}}$. Under these changes of variables, the equations \eqref{EF1}-\eqref{EF3} become
\begin{align}
  \partial_tD+ \text{div}(D\bv)&= 0\,, \label{EFb1}\\
  MD(\partial_t+\bv\cdot\nabla)\bv &= \hbar D\bv \times\nabla\times\bLambda - 
  D\nabla a-\frac{2}{\hbar}\nabla\boldsymbol{b}\cdot\tilde{\boldsymbol{s}}
+{  
  M^{-1}\partial_j(\tilde{\boldsymbol{s}}\cdot\nabla(D^{-1}\partial_j\tilde{\boldsymbol{s}}))}\,,\label{EFb2}\\
 \hbar M(\partial_t\tilde{\boldsymbol{s}}  + \text{div}(\bv\tilde{\boldsymbol{s}} ))&= \tilde{\boldsymbol{s}} 
 \times \Big(\hbar\text{div}(D^{-1}\nabla\tilde{\boldsymbol{s}} ) - {2M}\boldsymbol{b}\Big) 
\label{EFb3}\,.
\end{align}
For example, in the case of the spin-boson model, one has $a(\br)=M\omega^2r^2/2$ and $\boldsymbol{b}(\br)=(\mathcal{D},0,\boldsymbol{C}\cdot\br)/2$,  where $\boldsymbol{C}$ and $\mathcal{D}$ are time-independent and spatially constant.

Then, equations \eqref{EFb1}-\eqref{EFb3} possess the Euler-Poincar\'e Lagrangian
\begin{align}\label{EFHDbLagr}
  \ell(\bv, D, \boldsymbol\Xi,\tilde{\boldsymbol{s}} ) = \int\! \left[\frac{1}{2}MD|\bv|^2 + 
  D(\hbar\bv\cdot\bLambda- a)+ \tilde{\boldsymbol{s}}\cdot\Big(\boldsymbol\Xi -\frac{2}{\hbar}\boldsymbol{b}\Big) 
  -\frac{|\nabla\tilde{\boldsymbol{s}}|^2}{2MD}\right] \text{d}^3r\,,
\end{align}
where $\Xi = -i\bXi\cdot\bsigma/2$ having used the Lie algebra isomorphism $(\mathfrak{su}(2),[\,\cdot\,,\,\cdot\,]) \cong (\mathbb{R}^3, \,\cdot\,\times\,\cdot\,)$ \cite{MarsdenRatiu2013}, with the variations \eqref{vars} as well as
\[
\delta\tilde{\boldsymbol{s}}= {\boldsymbol\Upsilon}\times\tilde{\boldsymbol{s}}- \operatorname{div}(\bw\tilde{\boldsymbol{s}})\,,\qquad \delta \bXi = \partial_t{\boldsymbol\Upsilon} + {\boldsymbol\Upsilon}\times\bXi - \bw\cdot\nabla\bXi + 
\bv\cdot\nabla{\boldsymbol\Upsilon}\,.
\]
Here, $\boldsymbol\Upsilon(\br,t)$ is arbitrary and vanishing at the endpoints. 
We notice that in using the variable $\tilde\bs:=D\bs$, the quantum potential 
term has been absorbed in the Lagrangian and correspondingly no longer appears 
explicitly in the hydrodynamic equation of motion \eqref{EFb2}.

Analogously, in the case of electronic two-level systems, the Lie-Poisson bracket \eqref{LPB-EF} becomes
\begin{align}
\begin{split}
\{f,g\}(\bm,D,\tilde\bs)=&\,\int\!\bm\cdot\left(\frac{\delta g}{\delta \bm}\cdot\nabla\frac{\delta f}{\delta \bm}-\frac{\delta f}{\delta \bm}\cdot\nabla\frac{\delta g}{\delta \bm}\right)\de ^3 r
\\
&\,
-
\int \!D\left(\frac{\delta f}{\delta \bm}\cdot\nabla\frac{\delta g}{\delta D}-\frac{\delta g}{\delta \bm}\cdot\nabla\frac{\delta f}{\delta D}\right)\de ^3 r
\\
&\,
-\int \tilde\bs\cdot\left(\frac{\delta f}{\delta \tilde\bs}\times\frac{\delta g}{\delta \tilde\bs}+\frac{\delta f}{\delta \bm}\cdot\nabla\frac{\delta g}{\delta \tilde\bs}-\frac{\delta g}{\delta \bm}\cdot\nabla\frac{\delta f}{\delta  \tilde\bs}\right)\de ^3 
r\,.
\end{split}\label{LPB-Pauli}
\end{align}
This is accompanied by the Hamiltonian
 \begin{align}
   h(D,\bm,\tilde\bs) &= \int \!\left(\frac{|\bm-\hbar D\bLambda|^2}{2MD} + \frac{|\nabla \tilde\bs|^2}{2MD} + \frac{2}{\hbar}\boldsymbol{b}\cdot\tilde\bs +Da \right)
{\rm d}^3r\,,
\end{align}
where we recall the definition $\bm:=MD\bv+\hbar D\bLambda$.

 \subsection{The Pauli equation with hydrodynamic vortices}\label{Sec:Pauli}
Having considered a spinless particle in a magnetic field in Section \ref{SchrodingerReconstruction}, here we broaden our treatment to include particles with spin. To do so, one must consider the Pauli equation which captures the interaction of the particle's spin with an external electromagnetic field. Firstly, to describe a spin 1/2 particle (of charge $q=1$, mass $M$), one considers the two-component spinor wavefunction  
\begin{align}
  \Psi(\br) = \begin{pmatrix}
  \Psi_1(\br)
 \\
 \Psi_2(\br)
  \end{pmatrix}\,,
\end{align}
where now $\Psi \in L^2(\mathbb{R}^3)\otimes\mathbb{C}^2$ and is normalized such that $\int \Psi^{\dagger}\Psi \,\text{d}^3r=1$. Then, the dynamics are 
given by the Pauli equation which amounts to the Schr{\"o}dinger equation $i\hbar\partial_t\Psi = \widehat{H}_{\text{tot}}\Psi$ 
with the Hamiltonian operator
\begin{align}
  \widehat{H}_{\text{tot}}=\frac{(-i\hbar\nabla-\bA)^2}{2M} - 
  \frac{\hbar}{2M}\bB\cdot\widehat{\boldsymbol{\sigma}} + V\mathbbm{1}\,,
\end{align}
in which $\bA(\br)$ is the constant magnetic potential, $\bB:=\nabla\times\bA$ is the magnetic field and $V(\br)$ is a scalar potential. 

In order to proceed with the method proposed in the previous section, we follow the idea outlined in \cite{Spera2016, BohmSchillerTiomnoA1955, BohmSchillerB1955} and decompose the spinor wavefunction into the form 
$  \Psi= \sqrt{D(\br,t)}\theta(\br,t)\ket{\phi(\br,t)}$ (here we use the Dirac notation convention for the `electronic 
factor'),
in which $\phi$ satisfies the partial normalization condition $\braket{\phi(\br)|\phi(\br)}=:\|\phi(\br)\|^2=1$. Clearly, we are in the situation described in Sections \ref{Subsec:Nonad} and \ref{SubSec:2Level}
 so that, upon absorbing the additional external magnetic field, the  
 fluid velocity \eqref{philip} now reads
\begin{align}
  \bv = M^{-1}(\bar\bnu + \boldsymbol{\cal A} - \bA)\label{PauliVelocity}\,,
\end{align}
and one can immediately specialize the results from the previous section, so that the Lagrangian \eqref{EFHDbLagr} becomes
\begin{multline}\label{EFHDbLagr2}
  \ell(\bv, D, \boldsymbol\Xi,\tilde{\boldsymbol{s}} ) = \int\! \bigg[\frac{1}{2}MD|\bv|^2 + 
  D\bv\cdot(\hbar\bLambda + \bA) - DV + \tilde{\boldsymbol{s}}\cdot\Big(\boldsymbol\Xi +M^{-1}\bB\Big) 
  -\frac{|\nabla\tilde{\boldsymbol{s}}|^2}{2MD}\bigg] \text{d}^3r\,.
\end{multline}
%
%
In turn, this produces the equations of motion
\begin{align}
  \partial_tD+ \text{div}(D\bv)&= 0\,, \label{PauliEOM1}\\
 \begin{split}\label{PauliEOM2} MD(\partial_t+\bv\cdot\nabla)\bv &=  D\bv \times\nabla\times (\hbar\bLambda + \bA)  - 
  D\nabla V \\
  &\qquad\qquad\qquad\qquad+ M^{-1}\nabla\bB\cdot\tilde{\boldsymbol{s}}
+{ 
  M^{-1}\partial_j(\tilde{\boldsymbol{s}}\cdot\nabla(D^{-1}\partial_j\tilde{\boldsymbol{s}}))}\,,
 \end{split}\\
 M(\partial_t\tilde{\boldsymbol{s}}  + \text{div}(\bv\tilde{\boldsymbol{s}} ))&= \tilde{\boldsymbol{s}} 
 \times \big(\text{div}(D^{-1}\nabla\tilde{\boldsymbol{s}} ) +\bB\big) 
\label{PauliEOM3}\,.
\end{align}
In the special case of $\bLambda = 0$, this agrees with the results of \cite{Takabayasi1955} 
upon changing variables back to $\bs=\tilde{\boldsymbol{s}}/D$. In more generality, the corresponding circulation theorem for 
a loop $c(t)$ moving with the fluid velocity $\bv$ reads
\begin{align}
\begin{split}
\frac{\de}{\de t}\oint_{c(t)} (M\bv + \hbar\bLambda + \bA)\cdot\de\br  &\\
=\frac{\de}{\de t}\oint_{c(t)} \boldsymbol{\cal A}\cdot\de\br &=
-\frac{1}{MD}\oint_{c(t)} \Big[\nabla\tilde{\boldsymbol{s}}\cdot \big(\bB + \text{div}(D^{-1}\nabla\tilde{\boldsymbol{s}})\big)\Big]\cdot\de\br\,,
\end{split}
\end{align}
which, as in \cite{FoHoTr19}, gives an expression for the time evolution of the 
Berry phase, here governed by the external magnetic field and spin degrees of 
freedom. 
\begin{remark}[Mermin-Ho relation and Takabayasi vector]
In general, the circulation around a fixed loop $c_0$ (with a surface $S_0$ such that its boundary defines the loop $\partial S_0 =: 
c_0$) reads
{\begin{align}\label{MHo}
  \oint_{c_0} \bv\cdot\de\br = \int_{S_0} \nabla\times\bv\cdot\de\boldsymbol{S} 
  = M^{-1}\int_{S_0} \left(\frac{\hbar}{2}\boldsymbol{T} -\hbar\nabla\times\bLambda- 
  \bB\right)\cdot\de\boldsymbol{S}\,,
\end{align}
where $\boldsymbol{T}:= \epsilon_{ijk}n_i\nabla n_j\times\nabla n_k$ and $\bn:=2\bs/\hbar$ is the Bloch vector field. The relation $\nabla\times\boldsymbol{\cal A}=\hbar\boldsymbol{T}/2$
underlying \eqref{MHo} first appeared in Takabayasi's  work \cite{Takabayasi1955} in 1955, hence later motivating the name {\em Takabayasi vector} \cite{Bialynicki-Birula1995Weyl}, 
and can more explicitly be written in components as
\begin{align}
\begin{split}
  T_c &= \epsilon_{ijk}\epsilon_{abc}n_i(\partial_an_j)(\partial_bn_k)\\
  &= 2i\epsilon_{abc}\Big(\braket{\partial_b\phi|\partial_a\phi}-\braket{\partial_a\phi|\partial_b\phi}\Big)=\frac{2}{\hbar}{\cal B}_c\,,
\end{split}
\end{align}
relying on the definition $n_i = \braket{\phi|\sigma_i\phi}=\phi^*_a(\sigma_i)_{ab}\phi_b$ and the property
\begin{align*}
  \epsilon_{ijk}(\sigma_i)_{ab}(\sigma_j)_{cd}(\sigma_k)_{ef} = 
  2i\Big(\delta_{af}\delta_{cb}\delta_{ed}-\delta_{ad}\delta_{cf}\delta_{eb}\Big)\,.
\end{align*}
Despite the earlier discovery by Takabayasi, the relation $2\hbar^{-1}\nabla\times\boldsymbol{\cal A}= \epsilon_{ijk}n_i\nabla n_j\times\nabla n_k$ is more commonly known as the {\rm Mermin-Ho relation} \cite{MerminHo1976} after its appearance in the context of superfluids two decades later. For applications of the Mermin-Ho relation in the more general context of complex fluids, see also \cite{Holm2002}. 
}
\end{remark}

In \cite{Takabayasi1983}, Takabayasi pointed out the existence of vortex structures in spin hydrodynamics starting from the analysis of the Pauli equation and his work was later revived in \cite{Bialynicki-Birula1995Weyl}, where it was extended to comprise the relativistic Weyl equation. Motivated by these old works and proceeding in analogy with the previous sections, here we shall discuss the Rasetti-Regge dynamics of hydrodynamic vortices upon extending the arguments in Section \ref{Sec:Vortices}.
To do so, we write the Rasetti-Regge hydrodynamic Lagrangian
\begin{align}
 \bar{\ell}(\bR,\partial_t{\bR},\bv, D, \boldsymbol\Xi,\tilde{\boldsymbol{s}}) = \frac{1}{3}\int\partial_t\bR\cdot\bR \times\bR_{\sigma} \,{\rm d}\sigma +   \ell(\bv,\bR, D, \boldsymbol\Xi,\tilde{\boldsymbol{s}} )\,,
\end{align}
where $\ell$ is obtained upon replacing \eqref{p-vort} in \eqref{EFHDbLagr2}. The resulting vortex equation of motion is $ \partial_t\bR = M(D\bv)_{\br=\bR}+\kappa\bR_\sigma$ (where as before $\kappa$ is an arbitrary 
quantity), complemented by the hydrodynamic form of the Pauli equations 
\eqref{PauliEOM1}-\eqref{PauliEOM3}, in which again $\bLambda$ is written in terms of the vortex filament according to \eqref{p-vort}.

\begin{remark}[Pauli equation with hydrodynamic vortices]
  Naturally, this above construction can also be applied to the full Pauli spinor 
  $\Psi(\br,t)$, in which case one writes the Rasetti-Regge Dirac-Frenkel Lagrangian 
\begin{multline}
L(\bR,\partial_t{\bR},\Psi,\partial_t\Psi) = \frac{1}{3}\int\partial_t\bR\cdot\bR \times\bR_{\sigma} \,{\rm d}\sigma \\+\operatorname{Re}\int\! \bigg(i\hbar\Psi^{\dagger}\partial_t\Psi
- \Psi^{\dagger} \bigg[\frac{(-i\hbar\nabla-(\hbar\bLambda + \bA))^2}{2M}-\frac{\hbar}{2M}\bB\cdot\widehat{\bsigma}+V\bigg]\Psi\bigg){\rm d}^3{r}  \,,
\end{multline}
so that the coupled system reads 
\begin{align}
\partial_t{\bR}=&\,{\hbar}\,\mathbb{P}\left(\operatorname{Im}(\Psi^{\dagger}\nabla\Psi)-|\Psi|^2(\bLambda+\hbar^{-1}\bA)\right)\!\big|_{\br=\bR\,}
+\kappa\bR_\sigma\label{PauliVortex2}\,,
\\
i\hbar\partial_t\Psi=&\,\frac1{2M}{(-i\hbar\nabla-(\hbar\bLambda+\bA))^{2}}\Psi-\frac{\hbar}{2M}\bB\cdot\widehat{\bsigma}\,\Psi+V\,\Psi\label{PauliRecon}\,.
\end{align}
Upon expanding $\Psi$ in terms of the exact factorization and recalling the definition of the velocity \eqref{PauliVelocity}, we see the agreement in the vortex equations. Upon performing a long calculation 
similar to appendix \ref{LongCalc}, one can also reconstruct the Pauli equation 
given here from the hydrodynamic equations \eqref{PauliEOM1}-\eqref{PauliEOM3}.
\end{remark}

\section{Non-Abelian generalizations}\label{Sec:Non-Abelian}
As mentioned in remark \ref{jack}, the QHD gauge connection $\bnu$ from Section \ref{Sec:PhaseConnections} was introduced by following an approach that invokes zero curvature, i.e. $\nabla\times\bnu=0$, although this relation was later relaxed to comprise a non-zero initial curvature. In this section, motivated by the appearance of the differential of the spin vector $\nabla\bs$ appearing in the previous sections, we shall include the possibility of non-Abelian groups, therefore extending our treatment to a whole class of models employing the general relation $\nabla n = -\gamma n$, where $n\in\mathcal{F}(\mathbb{R}^3,M)$ is an order parameter field and $\gamma$ is a gauge connection corresponding to an arbitrary gauge group $\mathcal{F}(\mathbb{R}^3,G)$, where $G$ acts on $M$. 
{Nevertheless, even though the relation $\nabla n = -\gamma n$ again implies zero curvature,  the equations resulting from Hamilton's principle still allow for a more general non-trivial connection whose curvature again arises as an initial condition. }

To begin, we consider the previous treatment of QHD. {We consider a ${U}(1)$ connection defined by $\nabla\theta = -\bnu\theta$, \eqref{NuConnectionDefinition}, and see that this definition immediately implies that the connection has zero 
curvature. Indeed, one has $0 = \nabla\times\nabla\theta=- \nabla\times(\bnu\theta)= - (\nabla\times\bnu)\theta$, so that $\nabla\times\bnu$=0.}
Now we consider the more general case of an order parameter $n \in \mathcal{F}(\mathbb{R}^3,M)$ (where at this point $M$ is an arbitrary manifold) whose evolution is given by an element $g$ of the Lie group 
$\mathcal{F}(\mathbb{R}^3,G)$,
\begin{align}
  {n}(\bx,t)={g}(\bx,t){n}_0(\bx)\,.
\end{align}
Then, one can construct a gauge connection $\gamma$ from the gradient of ${n}$ as 
follows,
\begin{align}
    \begin{split}
  \nabla{n} &= \nabla {g} {n}_0 + {g}\nabla{n}_0\\
  &= \nabla {g} \,{g}^{-1} {n} - {g} \gamma_0{n}_0\\
  &= - (-\nabla {g}\,{g}^{-1}+ {g}\gamma_0{g}^{-1}){n} := -\gamma{n}\,,
  \end{split}
\end{align}
and we have $\nabla{n} = -\gamma{n}$. Here, we show that such a 
relation must mean that the connection is trivial, i.e. has zero curvature: $\Omega_{ij}=\partial_i\gamma_j-\partial_j\gamma_i+[\gamma_i,\gamma_j]$.
Writing our defining relation in terms of differential forms as $\text{d}{n}=-\gamma{n}$, we compute the 
following:
\begin{align}
    \begin{split}
  0= \text{d}^2{n} &=  -\text{d}(\gamma_j{n}\,\text{d}x^j)\\
  &= -\Big((\partial_{[i}\gamma_{j]}){n} + \gamma_{[j}(\partial_{i]}{n})\Big)\,\text{d}x^i\wedge \text{d}x^j\\
    &= -\Big(\partial_{[i}\gamma_{j]} - \gamma_{[j}\gamma_{i]}\Big){n}\,\text{d}x^i\wedge \text{d}x^j\\
        &= -\frac{1}{2}\Big(\partial_i\gamma_j - \partial_j\gamma_i + \gamma_i\gamma_j - \gamma_j\gamma_i\Big){n}\,\text{d}x^i\wedge \text{d}x^j\\
                &= -\frac{1}{2}\Omega_{ij}{n}\,\text{d}x^i\wedge \text{d}x^j\\
                &{= -\Omega{n}}
                \,,
  \end{split}
\end{align}
where the square brackets denote  index anti-symmetrization.
At this point, we observe that since $\gamma := -\nabla {g}\,{g}^{-1}+ {g}\gamma_0{g}^{-1}$, then 
\[
\Omega=g\Omega_0 g^{-1}
\]
and hence it follows that $0= \Omega_0 n_0$. Thus, we end up in a situation in which either $\Omega_0=0$ thus rendering $\Omega={\rm d}^\gamma\gamma=0$ for all time, or $n_0$ belongs to the kernel of $\Omega_0$, a statement that we cannot impose in general. 
As a specific example of this relation, reconsider the material involving the spin vector in Section \ref{Sec:ExactFact}. There, the order parameter is the spin vector $\bs \in \mathcal{F}(\mathbb{R}^3,\mathbb{R}^3)$ which evolves under the action of the rotation group $G=SO(3)$ so that $g=R(\bx,t)$ and the above formula 
specialize to
\begin{align}
  \bs(\bx,t)&=R(\bx,t)\bs_0(\bx)\,,\\
  \nabla\bs &= -\widehat{\gamma}\bs = -\bgamma \times \bs\,,\\
  0 &= \widehat{\Omega}\bs = \bOmega \times \bs\,,
\end{align}
having used the Lie algebra isomorphism ({\it hat map}) 
\begin{align*}
(\mathfrak{so}(3),[\,\cdot\,,\,\cdot\,])\cong(\mathbb{R}^3,\,\cdot\,\times\,\cdot\,)\,.
\end{align*}
Despite this result, we now turn our attention to the particular step in which the gauge connection is introduced in the Lagrangian of a field theory. As an example, consider a Lagrangian of the general type
\begin{align*}
    \ell=\ell(n,\dot{n},\nabla n)\,,
\end{align*}
where $n \in \mathcal{F}(\mathbb{R}^3, M)$. For example,  in the case of the Ericksen-Leslie theory of  liquid crystal nematodynamics, we have  $M=S^2$ \cite{Gay-BalmazRatiuTronci2012,Gay-BalmazRatiuTronci2013}. According to the previous discussion one can let $n$ evolve under local rotations  $g\in \mathcal{F}(\mathbb{R}^3, G)$, so that $n(t)=g(t)n_0$. This leads to introducing a gauge connection such that $\nabla n = -\gamma n$. In turn, the latter relation can be used to obtain a new Lagrangian of the form $\ell=\ell(n, \xi, \gamma)$, where $\xi:=\dot{g}g^{-1}$. Then, the Hamilton's principle associated to this new Lagrangian produces a more general set of equations in which $\gamma$ is allowed to have a non-zero curvature (constant if the gauge group is Abelian), as it appears by taking the covariant differential ${\rm d}^\gamma=\de + \gamma$ of the evolution equation $\partial_t\gamma+{\rm d}\xi=[\xi,\gamma]$ thereby producing $\partial_t\Omega=[\xi,\Omega]$. Notice that, while one has $(\partial_t-\xi)({\rm d}n+\gamma n)=0$, allowing for a non-trivial connection enforces the presence of an $M-$valued one-form $\phi(t)=g(t)\phi_0\in\Omega^1(\mathbb{R}^3,M) $ such that ${\rm d}n(t)+\gamma(t) n(t)=\phi(t)\neq0$. This observation (basically amounting to $\nabla n_0\neq-\gamma_0 n_0$) offers a general method for constructing defect theories whose defect topology does not depend on time. This is precisely the approach that we followed in Section \ref{Sec:PhaseConnections} and applied in the following sections.

\section{Conclusions}
This paper has introduced a new method for introducing holonomy in quantum hydrodynamics. Upon focusing on single-valued phase-factors rather than multi-valued phases, we have shown how a constant non-zero curvature can be naturally included to incorporate a non-trivial geometric phase in Madelung's equations. Also, it was shown how this method corresponds to simply applying minimal coupling at the level of Schr\"odinger's equation. In turn, this new picture led to the possibility of dealing with vortex singularities in the hydrodynamic vorticity. While topological singularities may be captured by the present treatment, our attention focused on vortex filaments of hydrodynamic type. By using the Rasetti-Regge framework, coupled equations were presented for the evolution of a Schr\"odinger wavefunction interacting with a hydrodynamic vortex filament.

As a first application of our approach, we considered the Born-Oppenheimer approximation in adiabatic molecular dynamics. After reviewing the variational setting of adiabatic dynamics, we presented the standard approach along with a modified approach presented in Appendix \ref{GaussianAppendix} and exploiting Gaussian wavepackets. Remarkably, in the latter approach, conical intersections are filtered by the Gaussian convolution kernel so that the nuclear motion occurs on a smoothed electron energy surface in agreement with the recent proposal by Gross and collaborators \cite{Min14, RequistEtAl2016}. A similar approach was then used on the variational side to incorporate vortex singularities in the Born-Oppenheimer approximation so that nuclei interact with a hydrodynamic vortex incorporating molecular geometric phase effects.

Last, the treatment was extended to the exact factorization of wavefunctions depending on more than one set of coordinates. Recently revived within the chemical physics community, this method has been widely used in the literature on the Pauli equation for a spin particle in an electromagnetic field. After reviewing the theory in both its variational and Hamiltonian variants, we specialized to consider the case of electronic two-level systems thereby studying the dynamics of  the spin  density vector. Finally, motivated by previous work by Takabayasi, we used this setting to include vortex filament dynamics in the quantum hydrodynamics with spin.

\addtocontents{toc}{\protect\setcounter{tocdepth}{0}}

\appendix

\section{Schr{\"o}dinger reconstruction calculation}\label{LongCalc}
This appendix presents the explicit calculations that reconstruct the Schr{\"o}dinger 
equation from the QHD equations in Section \ref{Sec:PhaseConnections}. We begin with the expansion
\begin{align*}
  i\hbar\partial_t\psi &= i\hbar(\partial_t R\,\theta + R\,\partial_t 
  \theta)\,.
\end{align*} 
Then, we find $\partial_t R$ from equation \eqref{transport} as
\begin{align*}
 \partial_t R &= - \nabla R \cdot \frac{\bar\bnu}{m} - 
    \frac{R}{2m}\text{div}(\bar\bnu)\,.
\end{align*}
Next, using equation \eqref{NuEvo} and \eqref{xiHamiltonJacobi1} we can also 
compute $\partial_t\theta$ and obtain
\begin{align*}
\partial_t\theta &= -\frac{i}{\hbar}\left(\frac{|\bar\bnu|^2}{2m}+V +V_Q 
\right)\theta\,.
\end{align*}
Hence, at this stage the Schr{\"o}dinger 
equation reads
\begin{align*}
i\hbar\partial_t\psi &= \left[-\frac{i\hbar}{m}\left(\frac{\nabla R}{R}\cdot\bar\bnu\right)-\frac{i\hbar}{2m}\text{div}(\bar\bnu)+\frac{|\bar\bnu|^2}{2m} +V_Q\right]\psi 
+ V\psi\,.
\end{align*}
Clearly, we must manipulate the kinetic term to get back to 
$\psi$. To do so, we recall the relations
\begin{align*}
  \frac{\nabla R}{R} = \frac{\text{Re}(\psi^*\nabla\psi)}{|\psi|^2}\,, 
  \qquad
    \bar\bnu =  \frac{\hbar\text{Im}(\psi^*\nabla\psi)}{|\psi|^2} -\hbar
{\bLambda}\,, \qquad
  V_Q=  - \frac{\hbar^2}{2m}\left(\frac{|\nabla R|^2}{R^2} + \text{div}\left(\frac{\nabla 
  R}{R}\right)\right)\,,
\end{align*}
and compute term-by-term. Firstly, 
\begin{align*}
  -\frac{i\hbar}{m}\left(\frac{\nabla R}{R}\cdot\bar\bnu\right)&= 
  -\frac{i\hbar^2}{m}\frac{\text{Re}(\psi^*\nabla\psi)\cdot\text{Im}(\psi^*\nabla\psi)}{|\psi|^2|\psi|^2}+ 
  \frac{i\hbar^2}{m}\frac{\text{Re}(\psi^*\nabla\psi)}{|\psi|^2}\cdot{\bLambda}\,.
\end{align*}
Secondly,
\begin{align*}
  -\frac{i\hbar}{2m}\text{div}(\bar\bnu)&= -\frac{i\hbar^2}{2m}\left(\nabla((\psi^*\psi)^{-1})\cdot\text{Im}(\psi^*\nabla\psi)+ 
\frac{\cancelto{0}{\text{Im}(\nabla\psi^*\cdot\nabla\psi)}}{|\psi|^2}+\frac{\text{Im}(\psi^*\Delta\psi)}{|\psi|^2}\right)\\
&= 
-\frac{i\hbar^2}{2m}\left(-(\psi^*\psi)^{-2}(\nabla\psi^*\psi+\psi^*\nabla\psi)\cdot\text{Im}(\psi^*\nabla\psi)+\frac{\text{Im}(\psi^*\Delta\psi)}{|\psi|^2}\right)\\
&= -\frac{i\hbar^2}{2m}\frac{\text{Im}(\psi^*\Delta\psi)}{|\psi|^2} + 
\frac{i\hbar^2}{m}\frac{\text{Re}(\psi^*\nabla\psi)\cdot\text{Im}(\psi^*\nabla\psi)}{|\psi|^2|\psi|^2}\,.
\end{align*}
where in the second line we have used that ${\bLambda} = 
-\nabla\times\bbeta$ so that its gradient vanishes. Thirdly,
\begin{align*}
  \frac{|\bar\bnu|^2}{2m} 
&= \frac{\hbar^2}{2m}\frac{\text{Im}(\psi^*\nabla\psi)\cdot\text{Im}(\psi^*\nabla\psi)}{|\psi|^2|\psi|^2} 
- 
\frac{\hbar^2}{m}\frac{\text{Im}(\psi^*\nabla\psi)}{|\psi|^2}\cdot{\bLambda}+\frac{\hbar^2}{2m}|{\bLambda}|^2\,.
\end{align*}
Finally,
\begin{align*}
  V_Q 
    &= -\frac{\hbar^2}{2m}\left(\frac{\text{Re}(\psi^*\nabla\psi)\cdot\text{Re}(\psi^*\nabla\psi)}{|\psi|^2|\psi|^2} +\frac{\text{Re}(\nabla\psi^*\cdot\nabla\psi)}{|\psi|^2}+\frac{\text{Re}(\psi^*\Delta\psi)}{|\psi|^2}+ 
  \nabla((\psi^*\psi)^{-1})\cdot\text{Re}(\psi^*\nabla\psi)\right)\\
    &= -\frac{\hbar^2}{2m}\left(\frac{\text{Re}(\psi^*\nabla\psi)\cdot\text{Re}(\psi^*\nabla\psi)}{|\psi|^2|\psi|^2} +\frac{|\nabla\psi|^2}{|\psi|^2}+\frac{\text{Re}(\psi^*\Delta\psi)}{|\psi|^2} 
  -2\frac{\text{Re}(\psi^*\nabla\psi)\cdot\text{Re}(\psi^*\nabla\psi)}{|\psi|^2|\psi|^2}\right)\\
   &= -\frac{\hbar^2}{2m}\left(-\frac{\text{Re}(\psi^*\nabla\psi)\cdot\text{Re}(\psi^*\nabla\psi)}{|\psi|^2|\psi|^2} +\frac{|\nabla\psi|^2}{|\psi|^2}+\frac{\text{Re}(\psi^*\Delta\psi)}{|\psi|^2} 
   \right)\,.
\end{align*}
All together the kinetic term reads
\begin{align*}
  -\frac{i\hbar}{m}\left(\frac{\nabla R}{R}\cdot\bar\bnu\right)-\frac{i\hbar}{2m}\text{div}(\bar\bnu)+\frac{|\bar\bnu|^2}{2m} 
  +V_Q&= -\frac{\hbar^2}{2m}\left(\frac{\text{Re}(\psi^*\Delta\psi)}{|\psi|^2}+ \frac{i\text{Im}(\psi^*\Delta\psi)}{|\psi|^2}\right) 
  \\
  &+ \frac{i\hbar^2}{m}\left(\frac{\text{Re}(\psi^*\nabla\psi)}{|\psi|^2}+ 
  \frac{i\text{Im}(\psi^*\nabla\psi)}{|\psi|^2}\right)\cdot{\bLambda}+\frac{\hbar^2}{2m}|{\bLambda}|^2\\
  &+\frac{\hbar^2}{2m}\frac{\text{Re}(\psi^*\nabla\psi)\cdot\text{Re}(\psi^*\nabla\psi)}{|\psi|^2|\psi|^2}
\\
&\qquad+\frac{\hbar^2}{2m}\frac{\text{Im}(\psi^*\nabla\psi)\cdot\text{Im}(\psi^*\nabla\psi)}{|\psi|^2|\psi|^2}- \frac{\hbar^2}{2m}\frac{|\nabla\psi|^2}{|\psi|^2}\,,
\end{align*}
at which point we rewrite the following terms:
\begin{align*}
  \frac{\hbar^2}{2m}\frac{\text{Re}(\psi^*\nabla\psi)\cdot\text{Re}(\psi^*\nabla\psi)}{|\psi|^2|\psi|^2}
+\frac{\hbar^2}{2m}\frac{\text{Im}(\psi^*\nabla\psi)\cdot\text{Im}(\psi^*\nabla\psi)}{|\psi|^2|\psi|^2}  
= \frac{\hbar^2}{2m}\frac{|\psi^*\nabla\psi|^2}{|\psi|^2|\psi|^2} =  \frac{\hbar^2}{2m}\frac{|\nabla\psi|^2}{|\psi|^2} 
\,,
\end{align*}
and after cancellations one is left with
\begin{align*}
    -\frac{i\hbar}{m}\left(\frac{\nabla R}{R}\cdot\bar\bnu\right)-\frac{i\hbar}{2m}\text{div}(\bar\bnu)+\frac{|\bar\bnu|^2}{2m} 
  +V_Q&=-\frac{\hbar^2}{2m}\frac{\Delta\psi}{\psi} +
  \frac{i\hbar^2}{m}\frac{\nabla\psi}{\psi}\cdot{\bLambda} + 
  \frac{\hbar^2}{2m}|{\bLambda} |^2\,.
\end{align*}
Multiplying by $\psi$ and factorizing returns the desired result.

 \section{Adiabatic dynamics with Gaussian wavepackets}\label{GaussianAppendix}
Whilst the main focus of this paper revolves around employing hydrodynamic descriptions of 
quantum mechanics, in this appendix we approach the adiabatic problem in quantum 
chemistry through the use of frozen Gaussian wavepackets at the level of the 
variational principle. 

In line with the adiabatic separation of nuclei and electrons, we model the nuclear wavefunction $\Omega(\br,t)$ via a Gaussian wavepacket, which corresponds to the following replacements in the standard Madelung transform $\Omega=\sqrt{D}e^{iS/\hbar}$,
\begin{align}
  D(\br,t)=D_0(\br-\bq(t))\,,\qquad S(\br,t) = \bp(t)\cdot(\br-\bq(t)/2)\,,
\end{align}
where $D_0$ is a Gaussian of constant width (frozen). This implies that $\nabla S = \bp$ so that the Born-Oppenheimer total energy, corresponding to the Lagrangian 
\eqref{BO-Lagr}, reads
\begin{multline}
  h = \int D_0(\br-\bq)\left(\frac{|\bp+\boldsymbol{\cal A}|^2}{2M} + \frac{\hbar^2}{8M}\frac{|\nabla D_0(\br-\bq)|^2}{D_0(\br-\bq)^2} + 
\epsilon (\phi,\nabla\phi)\right)\,\rm{d}^3r\\
= \int D_0(\br-\bq)\left(\frac{|\bp+\boldsymbol{\cal A}|^2}{2M}  + 
\epsilon (\phi,\nabla\phi)\right)\,{\rm d}^3r + \text{const.}
\end{multline}
where we have noticed that the quantum potential term collapses to an irrelevant constant. 
At this stage, we invoke the commonly used approximation neglecting the second 
order coupling $\hbar^2\|\nabla\phi\|^2/(2M)$ in \eqref{effectivepotential} so that upon expanding the effective potential 
the total energy is
\begin{align}
  h = \frac{|\bp|^2}{2M} + M^{-1}\bp\cdot\, \overline{\!\boldsymbol{\cal A}}\, +\, \overline{\!E}\,,
\end{align}
where we have defined
\[
\overline{\!\boldsymbol{\cal A}}(\bq)=\int \!D_0(\br-\bq)\boldsymbol{\cal A}(\br)\,\de^3 r
\,,\qquad\qquad
\overline{\!E}(\bq)=\int \!D_0(\br-\bq) E(\br)\,\de^3 r
\,.
\]
Then, upon performing the Legendre transform $M\dot{\bq}= \bp+\,\overline{\!\boldsymbol{\cal A}}(\bq)$, one obtains
\begin{align}
  L(\bq,\dot{\bq})= \frac{M}{2}|\dot{\bq}|^2 - \dot\bq\cdot\,\overline{\!\boldsymbol{\cal A}}(\bq) 
   - \bar\epsilon(\bq)\,,
  \end{align}
  where we have defined
  \[
\bar\epsilon(\bq):=  \, \overline{\!E}(\bq)- \frac{1}{2M} \big|\,\overline{\!\boldsymbol{\cal A}}(\bq)\big|^2
\,.
  \]
Since both the energy surface and the Berry connection are smoothened by a Gaussian convolution filter, we notice that  the resulting equation of motion
\[
M\ddot{\bq}=-\dot\bq\times\nabla_{\!\bq}\times\,\overline{\!\boldsymbol{\cal A}}(\bq)-\nabla_{\!\bq\,}\bar\epsilon(\bq)
\]
is entirely regularized so that conical singularities are smoothened by the Gaussian convolution.

\subsection*{Acknowledgements}
Much of this work arises from our discussions during the MSRI research program ''Hamiltonian systems, from topology to applications through analysis" held from August 13, 2018 to December 14, 2018. We are grateful to the members of the Organizing Committee and to the MSRI staff at Berkeley for running the events of the program so cheerfully and efficiently. Also, we are grateful to Denys Bondar and Darryl Holm for several stimulating discussions on these and related topics. This material is partially based upon work supported by the NSF Grant No. DMS-1440140 while the authors were in residence at the MSRI, during the fall of 2018. In addition,
CT  acknowledges support from the Alexander von Humboldt Foundation (Humboldt Research Fellowship for Experienced Researchers) as well as from the German Federal Ministry for Education and Research.
MF acknowledges the Engineering and Physical Sciences Research Council (EPSRC) studentship grant 
EP/M508160/1. At the last revision stage, CT also benefited form the support of the John Templeton
Foundation Grant 62210. The opinions expressed in this publication are those of the authors and do not
necessarily reflect the views of the John Templeton Foundation.

\end{document}